\documentclass[preprint,12pt,authoryear]{elsarticle}

\usepackage{amssymb,amsfonts,amsmath}
\usepackage{algorithm}
\usepackage{algorithmicx}
\usepackage{algpseudocode}
\usepackage{booktabs}
\usepackage{color}
\usepackage{float}
\usepackage{bbding}
\usepackage{multicol,multirow}
\usepackage{inputenc}
\usepackage{subfig}
\usepackage{wrapfig}
\usepackage{url}
\usepackage{threeparttable}
\usepackage{nicefrac} 
\usepackage[table]{xcolor}
\usepackage{graphicx}
\usepackage{caption}
\usepackage[shortlabels]{enumitem}
\usepackage{xr-hyper}
\usepackage{verbatim}
\usepackage{wrapfig}
\usepackage{xspace}

\makeatletter
\DeclareRobustCommand\onedot{\futurelet\@let@token\@onedot}
\def\@onedot{\ifx\@let@token.\else.\null\fi\xspace}

\def\eg{\emph{e.g}\onedot} 
\def\ie{\emph{i.e}\onedot} 
 
\def\etc{\emph{etc}\onedot}

\makeatother

\journal{ISPRS Journal of Photogrammetry and Remote Sensing}

\begin{document}

\begin{frontmatter}



\title{ET-SDE: Efficient Terrain Stochastic Differential Equations for Multipurpose Digital Elevation Model Restoration}


\author[label1]{Tongtong Zhang}
\author[label1]{Yuanxiang Li*}
\author[label1]{Zongcheng Zuo}
\author[label2,label3]{Fan Mo}

\affiliation[label1]{organization={School of Aeronautics and Astronautics, Shanghai Jiao Tong University},
addressline={800 Dongchuan RD. Minhang District}, 
city={Shanghai},
postcode={200240}, 
country={China}}
\affiliation[label2]{organization={Land Satellite Remote Sensing Application Center},
city={Beijing},
postcode={100048}, 
country={China}}
\affiliation[label3]{organization={China-ASEAN Satellite Remote Sensing Applications,Ministry of Natural Resources of the People's Republic of China},
city={Nanning},
postcode={530221}, 
country={China}}

\begin{abstract}
Digital Elevation Models (DEMs) are indispensable in the fields of remote sensing and photogrammetry, with their refinement and enhancement being critical for a diverse array of applications.
Numerous methods have been developed for enhancing DEMs, but most of them concentrate on tackling specific tasks individually.
This paper presents a unified generative model for multipurpose DEM restoration, diverging from the conventional approach that typically targets isolated tasks. 
We modify the mean-reverting stochastic differential equation, to generally refine the DEMs by conditioning on the learned terrain priors.
The proposed Efficient Terrain Stochastic Differential Equation (ET-SDE) models DEM degradation through SDE progression and restores it via a simulated reversal process. Leveraging efficient submodules with lightweight channel attention, this adapted SDE boosts DEM quality and streamlines the training process.
The experiments show that
ET-SDE achieves highly competitive restoration performance on super-resolution, void filling, denoising, and their combinations, compared to the state-of-the-art work.
In addition to its restoration capabilities, ET-SDE also demonstrates faster inference speeds and the capacity to generalize across various tasks, particularly for larger patches of DEMs.
\end{abstract}





\begin{keyword}


digital elevation model,
stochastic differential equation,
diffusion probabilistic model, 
super-resolution,
void-filling,
denoising
\end{keyword}

\end{frontmatter}



\section{Introduction}
In recent years, there has been a growing demand for high-quality geospatial data across various scientific fields. This demand is driven by the increasing use of geospatial data in applications such as environmental monitoring, urban planning, and natural resource management \citep{dem_assess_review,dem_accuracy_review}. 
Digital Elevation Models (DEMs) are fundamental components in geospatial analysis, providing critical information about the Earth's topography, water resource management, and hydrological modeling \citep{hydrology,dem_landslides}. 
DEM can be acquired from different measurement technologies, such as field surveys, optical remote sensors, LiDAR, and InSAR.

Although the advancement of sensors and algorithms is capable of generating extensive DEMs covering large geographical areas, the limited precision of these measurement instruments significantly affects the availability of high-resolution DEMs, which are essential for terrain analysis. Additionally, terrain features exhibit multiscale characteristics \citep{dem_terrain,sota_dem_review}, and the scale conversion process can present challenges.
Moreover, the DEM production techniques and the acquisition configurations inevitably introduce systematic errors and random noises \citep{dem_assess_review}.
 In addition to noise, DEMs derived from stereo matching frequently contend with voids.
For instance, Shuttle Radar Topography Mission (SRTM) \citep{srtm_app}, Advanced Spaceborne
Thermal Emission and Reflectance Radiometer Global Digital Elevation Model (ASTER GDEM) \citep{aster},
and TerraSAR-X add-on for Digital Elevation Measurement (TanDEM-X) \citep{tandem} are affected by a large number of voids \citep{srtm}, which need to be reconstructed before the DEMs are used in further applications.

To refine DEM from low-resolution and voids, 
early attempts focused on interpolation techniques to estimate intermediate elevations between existing data points, including inverse-distance-weighted \citep{IDW}, natural nearest neighbor interpolation \citep{naturalneighbor}, spline interpolation \citep{comp_inter17}, \etc. 
Later, geostatistical information brings new inspirations, such as Kriging \citep{kriging} and its variants.
However, these methods could not capture fine-scale details and often resulted in artifacts \citep{comp_inter17,comp-inter19}.

For the DEM super-resolution task, with the progression of deep learning techniques in single-image super-resolution, an increasing number of studies have begun to tackle DEM enhancement by employing models that parallel those utilized in image super-resolution.
Several studies have arisen to apply high-performing image Super-Resolution models to DEM SR.
Convolutional Neural Networks (CNNs) were first employed in DEM Super-Resolution in D-SRCNN \citep{cnn1st_dem}, incorporating the architecture of SRCNN and facilitating the transition from image SR to DEM SR.
 \citep{xu2015nonlocal} introduced an initial nonlocal algorithm incorporating high-frequency information from learning examples, demonstrating superior outcomes compared to interpolation-based methods.  Another contribution \citep{gradient} presented a deep gradient prior network for DEM SR, leveraging the EDSR network architecture and integrating gradient loss in the training process. Later, multiscale supervision \citep{multiscale}, recursive feature extractor \citep{rspcn}, \etc have further enhanced the network representation capability. To better adapt to terrain properties, \citep{ridgeline} further utilizes fused topological information as supervision, and \citep{terrainfeatureaware} adaptively optimizes the feature-extracting module via deformable convolution.
With the development of generative image super-resolution models, generative adversarial networks (GANs) are progressively introduced to the DEM super-resolution domain. D-SRGAN \citep{d-srgan}, 

In the realm of void-filling, the evolution of the Conditional Generative Adversarial Network (CGAN) has prompted the adaptation of interpolation-based methods to integrate CGAN-based techniques for filling voids \citep{gan_interpolation,gan_interpolation2}.
Researchers have used CGAN models to fill DEM voids by utilizing the DEM features \citep{dem_void_gan}. Further advancements have included the incorporation of attention mechanisms \citep{dem_void_attgan} and the use of restricted topographic knowledge \citep{TKCGAN} to enhance the void-filling process.
More recently, a method called Diff-DEM has been proposed, which refers to the use of Denoising Diffusion Probabilistic Models to inpaint the voids in the DEM data.

For the denoising task, explicit error models can be utilized when extra information is provided \citep{dem_assess_review}. Later works \citep{coastaldem} adopt neural networks to improve horizontal resolution DEMs by employing a more comprehensive selection of inputs and far
larger training and testing sets. \citep{sar-speckle-dem} adopted CNN to remove speckle noise from DEMs generated from speckled SAR images.

However, the GAN-based model is vulnerable to mode corruption and unstable optimization \citep{sisr_review, gan_review}. The GAN-based models handle degradation removal and detail refinement simultaneously for the entire image grid. In contrast, diffusion models have demonstrated notable success in image restoration, exhibiting finer details and a more straightforward training process \citep{diffusion_review}. This novel branch of generative models, diffusion models, has recently yielded significant advancements in visual generation and restoration tasks.  
Besides, none of the works provide a compositive solution to different defects of DEMs.

This paper proposes a DEM refinement framework Efficient Terrain Stochastic Differential Equations (ET-SDE), for multipurpose restoration of DEM from different negative problems.
ET-SDE fully leverages a mean-reverting Stochastic Differential Equation conditioned on terrain features learned by deformable convolution.
With the unified representation of diffusion probabilistic models \citep{diffusion_review}, ET-SDE gradually diffuses images towards a pure noise distribution with an SDE, and then generates samples by learning and simulating the corresponding reverse-time SDE.
To adapt to the terrain model, ET-SDE is based on the mean-reverting SDE with a closed-form solution and adapted feature-extracting modules.
 The main contributions include:
 \begin{itemize}
     \item We propose ET-SDE, an efficient multipurpose DEM refinement framework ET-SDE for DEM super-resolution, void-filling, and denoising.
     \item We adapt the SDE model to enhance the performance by incorporating the terrain knowledge in both the pipeline and the losses.
     \item The ET-SDE yields pioneering results for each task, and has faster inference speed and generalization ability.
 \end{itemize}

The remainder of this paper is organized as follows. Section~\ref{sec:rw} gives a literature review of the related work in respective tasks. Section~\ref{sec:methods} gives the theory and the lightweight pipeline design. Section~\ref{sec:exp} gives experiments for respective tasks and the analysis. Section~\ref{sec:conclusion} contains the conclusion and future directions.
\section{Related work}
\label{sec:rw}
\subsection{Diffusion models for image restoration} 
Image restoration is the general task of restoring a high-quality image from a degraded, low-quality version. With diffusion models  emerging as a new branch of generative models,  
breakthroughs are made for image restoration \citep{diffusion_ir_review}.
The diffusion model transforms the complicated and unstable generation
process into several independent and stable reverse processes via Markov Chain modeling \citep{diffusion_review}. 
The three widely utilized models including NCSNs
Noise Conditioned Score Networks 
\citep{NCSNs},  
Denoising Diffusion Probabilistic Model  \citep{ddpm}, 
score matching stochastic differential equations \citep{score-sde}.
Diffusion models are widely applied to image restoration tasks. For image super-resolution, SR3 \citep{sr3} uses a typical DDPM framework with Unet. The following works applied different conditions, such as low-quality reference image \citep{palette}, pre-processed references \citep{edsr}, or revising diffusion process \citep{ir-sde,refusion}.
\citep{ilvr} uses an unconditional diffusion model to enable the
training-free conditional generation for image SR and image
translation.
\subsection{Enhancing the DEM resolution}
The native spatial resolution of DEMs is often limited by the sensor technology used for data acquisition, leading to the necessity for super-resolution techniques to enhance their quality. Early attempts focused on techniques such as bicubic interpolation to estimate intermediate elevations between existing data points. However, these methods could not capture fine-scale details and often resulted in artifacts.
In recent years, machine learning approaches have gained prominence in DEM super-resolution(DEM-SR). CNNs and GANs have shown promise in learning complex relationships within DEM data and generating high-resolution counterparts. Integrating machine learning techniques has significantly enhanced the performance of the state-of-the-art (SOTA) in DEM-SR. Enhanced Bilateral Filtering-based Continuous DEM (EBCF-CDEM) \citep{continuous} introduces the neural implicit representation to the DEM-SR task \citep{liif}.
\subsection{Restoration of DEM from voids and noises}
The quality of commonly
used DEM products, such as SRTM \citep{srtm_app}, ASTER GDEM \citep{aster},
and TanDEM-X \citep{tandem} are affected by a large number of voids \citep{srtm}, which need to be reconstructed before the DEMs are used in further applications.
Early void-filling methods primarily relied on simple interpolation techniques, such as bilinear or bicubic interpolation.
  The geostatistical interpolation method kriging \citep{kriging} models the spatial correlation of elevation values, where Ordinary kriging and universal kriging have been applied to capture the spatial variability of elevation data.
Spatial interpolation methods, including inverse distance weighted interpolation \citep{IDW}, and natural neighbor interpolation \citep{naturalneighbor}, remain fundamental in void-filling processes. These techniques estimate elevation values based on the known values in the surrounding neighborhood.
In recent years, with the development of GAN, voids filling methods based on interpolation have been adapted with CGAN \citep{gan_interpolation,gan_interpolation2}. \citep{dem_void_gan} filled DEM void with a CGAN by utilizing the DEM features, \citep{dem_void_attgan} further include attention mechanism, and \citep{TKCGAN} include restricted topographic knowledge.

As for the noises in DEMs, if auxiliary information is provided by other sensors, errors introduced solely by a single source can be removed with an explicit physical model, such as removing the vegetative land cover via known vegetation cover indices \citep{o2016multi}. 

\section{Methodology}
\label{sec:methods}
\subsection{Preliminaries}
\begin{figure}[H]
    \centering
    \includegraphics[width=13cm]{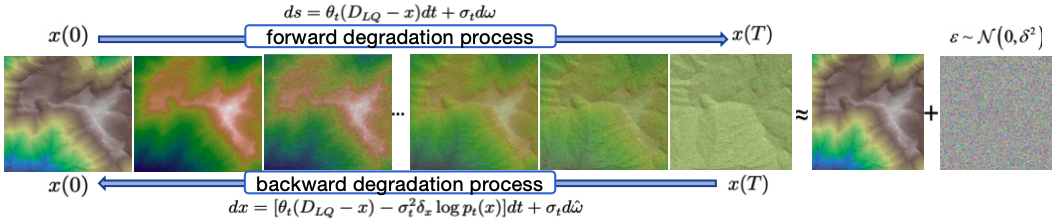}
    \caption{Overview of the ET-SDE to restore DEM from $D_{LQ}$ low resolution and irregular voids.
    The forward process of ET-SDE serves as a degradation from a high-quality DEM $x(0)=D_{HQ}$ to its low-quality counterpart $x(T) = D_{LQ}$ by progressively adding estimated noise. Inversely, recovering $D_{HQ}$ is obtained by simulating the reverse-time process.}
    \label{fig:sde}
\end{figure}

The score-based generative diffusion models, consisting of a forward process and a reverse process, can be represented by one SDE with its forward process and reverse-time process respectively. The SDE transforms the prior distribution back into the data distribution by slowly removing the noise \citep{sde}.
This paper adapts the mean-reverting SDE with a closed-form solution for unified DEM restoration shown in Fig.~\ref{fig:sde}.

\paragraph{\textbf{The forward process}}
In the forward process, noise is progressively added to the DEM until it becomes Gaussian noise.  
With $p_0$ being the initial distribution of the DEM data, $t\in [0, T]$ denotes the continuous time variable. The degeneration process, including resolution decrease, and void simulation, is considered a diffusion process in the form of SDE:
\begin{equation}
    dx = f(x, t)\mathrm dt + g(t)d\omega, x(0)\sim p_0(x),
    \label{eq:sde_forward}
\end{equation}
where $f(\cdot)$ is the drift function, $g(\cdot)$ is the dispersion function, and $\omega$ is a standard Wiener process, with $x(0)\in \mathbb{R}^d$ being an initial condition and $x(T)\in \mathbb{R}^d$ being the final state.
ET-SDE learns to gradually transform the data distribution into a fixed Gaussian noise, and then restore the high-quality DEM using the inverse process.

Given the input pairs $[D_{LQ}, D_{HQ}]$, to enable a tractable solution, Eq.~\ref{eq:sde_forward} is then transformed to the simplified version:
\begin{equation}
  dx = \theta_t(D_{LQ} - x)\mathrm dt + \sigma_t d\omega  
  \label{eq:deg_sde_forward}
\end{equation}
where $\theta_t$ indicates the speed of the mean-reversion, and $\sigma_t$ indicates the stochastic volatility.

To provide Eq.~\ref{eq:deg_sde_forward} with a closed-form solution, the SDE coefficients are set with the relation $\sigma^2_t/\theta_t = 2 \lambda^2$ for all times $t$. As \citep{ir-sde} proves, 
the marginal distribution $p_t(x)$ of $x(t)$ is derived as:
\begin{align}
    p_t(x) = \mathcal{N}(x(t)|m_t(x), v_t), \nonumber\\
    m_t(x) := D_{LQ} + (x(0)-D_{LQ})e^{-\bar{\theta_t}} \nonumber \\
    v_t := \lambda^2(1 - e^{-2\bar{\theta}_t})\label{eq:marginl_dist}
\end{align}
The mean $m_t(x)$ converges to $D_{LQ}$ and the variance $v_t$ converges to $\lambda^2$ when $t \rightarrow \infty$. 

\paragraph{\textbf{The reverse process}}
For the reverse process of the 
SDE in Eq.\ref{eq:sde_forward}, the DEM is reconstructed via its reverse-time representation:
\begin{equation}
dx = [f(x,t)-g(t)^2\nabla_x\log p_t(x)]\mathrm dt + g(t)d\hat{\omega}
    \label{eq:sde_backward}
\end{equation}

Specifically for the DEM super-resolution task, the analytically tractable version of the reverse SDE turns to:
\begin{equation}
    dx = [\theta_t(D_{LQ} - x) - \sigma_t^2\nabla_x \log p_t(x)]\mathrm dt + \sigma_td\hat{\omega}
\end{equation}

During the inference phase, according to Eq.\ref{eq:marginl_dist}, the state is sampled with  $x(t) = m_t(x)(x) + \sqrt{v_t}\epsilon_t, \epsilon_t \sim \mathcal{N}(0, I)$,
then the ground truth score function is simplified as:
\begin{align}
    \nabla_x \log p_t(x) & = -\frac{x(t) - m_t(x)}{v_t}\nonumber\\
    & := -\frac{\epsilon_t}{v_t}
    \label{eq:score_function}
\end{align}
where $\epsilon_t$ is the noise estimated by the neural network.
Similar to IR-SDE \citep{ir-sde}, we use a Unet-like architecture to estimate noise. To infer a high-quality DEM, we simulate the backward process with Eq.~\ref{eq:sde_backward}.
\begin{algorithm}[H]
\caption{Training of ET-SDE.}
\label{alg:train}
\begin{algorithmic}[1]
\Statex \textbf{INPUT} The degraded DEM patch $v=D_{LQ}$, its upsampled version $\mu=D_{LQ}$, and the high quality DEM patch $x_0=D_{HQ}$ 
\Statex \textbf{INIT} Random sample $\epsilon_t \approx \mathcal{N}(0, \sigma^2)$, $t\in [0, T]$.
\Statex \textbf{repeat} 

feat = TPE($v$, $\hat{\epsilon}_t$) + conv([$\mu$, $\epsilon_t$]); 

 // \textit{Terrain\quad Prior\quad Encoding}

 $\bar{\epsilon}_t = F_{\phi_{NN}}(feat, t)$

 // \textit{Noise Prediction with Network $F_{\phi_{NN}}$}

 $dx = [\theta_t(\mu-x)-g(t)^2 (-\frac{\bar\epsilon_t}{v_t})]\mathrm dt + g(t)d\hat{\omega}$

// \textit{Approximation of score for Eq.~\ref{eq:deg_sde_forward}.}

$L_{sde}(\Phi_{NN}) = \sum\limits_{t=0}^T\gamma_t\mathrm{E}[\|\hat{\epsilon}_{\Phi}(h_t, D_{LQ}, t) - \epsilon_t\|];$

// \textit{Calculate SDE loss in Eq.~\ref{eq:loss_sde}.}

$L = L_{sde} + L_{grad}$, $\nabla_{\Phi_{NN}} L$;

// \textit{Update network parameters with gradient descent.}
\Statex \textbf{until} converging
\end{algorithmic}
\end{algorithm}

\subsection{Pipeline}
\begin{figure}[t]
    \centering
    \includegraphics[width=14cm]{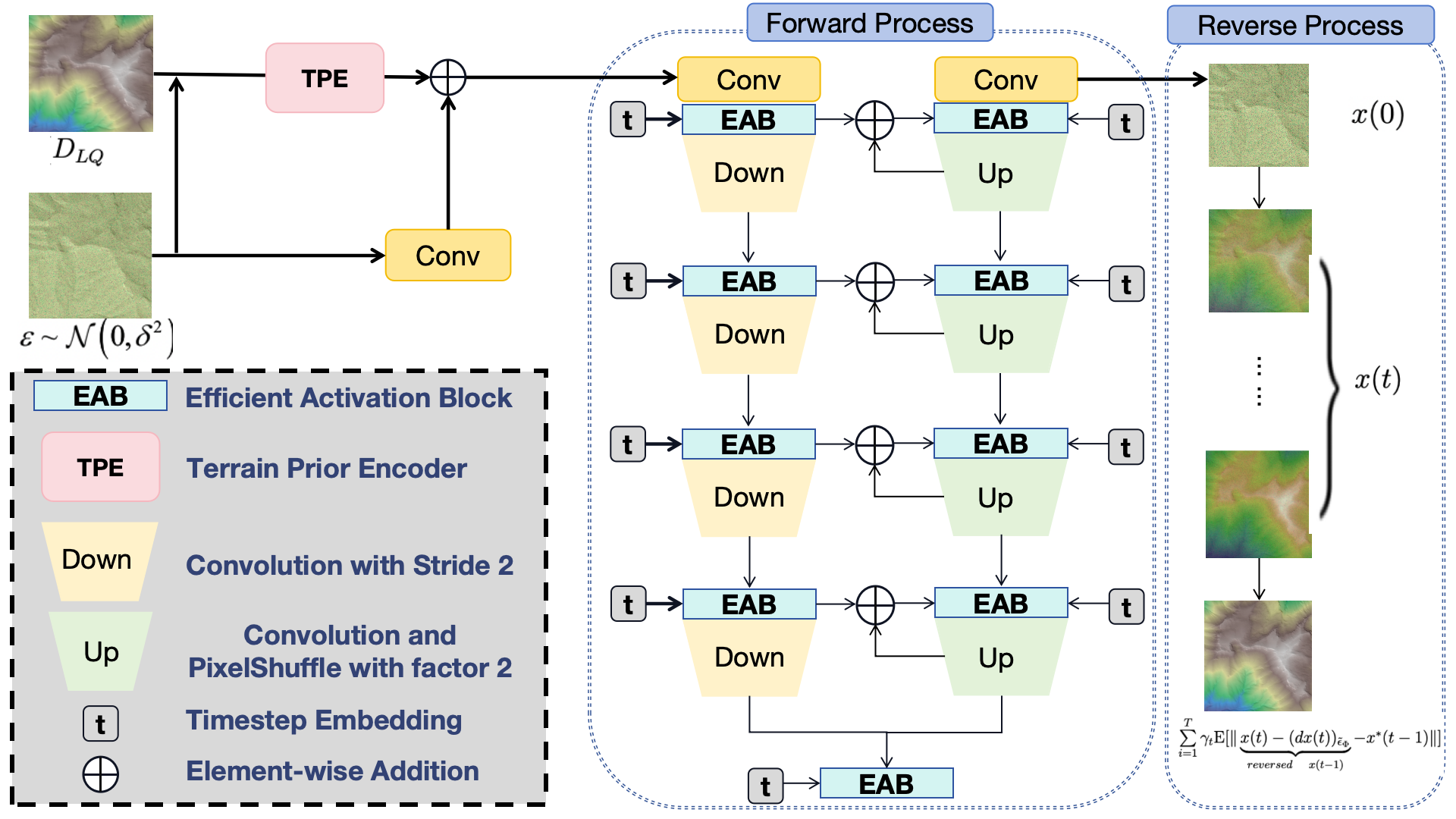}
    \caption{The Unet-shaped pipeline of ET-SDE is composed of a Terrain Prior Encoder (TPE), and several efficient submodules. The noise predictor $F_{\Phi_{NN}}$ with parameters $\Phi_{NN}$ is optimized in the forward process in Algorithm~\ref{alg:train}. }
    \label{fig:pipeline}
\end{figure}

Fig.~\ref{fig:pipeline} illustrates the entire pipeline. The noise predictor utilizes a modified architecture based on Unet-shaped architecture, with adjustments made in three different aspects:
\begin{itemize}
    \item To better capture the terrain-specific features rather than images, firstly the DEM is fed to the terrain prior encoder (TPE). The details are explained in Section \ref{sec:TPE}
    \item To enhance the efficiency of the noise predictor, the convolution blocks of the canonical U-Net are represented by the Efficient Activation Block (EAB), which is detailed in Section \ref{sec:eab}.
    \item The loss functions are specifically adapted to terrain features, which is detailed in Section~\ref{sec:loss} 
\end{itemize}
The training and inference (sampling) algorithms are given in Algorithm \ref{alg:train} and Algorithm \ref{alg:test} respectively.

\begin{algorithm}[H]
\caption{Inference of ET-SDE.}
\label{alg:test}
\begin{algorithmic}
\Statex \textbf{INPUT} The degraded DEM patch $v=D_{LQ}$, its upsampled version $\mu=D_{LQ}$, total step $T$. 
\Statex \textbf{OUTPUT} The restored DEM $\hat{D}_{HQ}$.
\Statex \textbf{INIT} Random sample $x_T\approx \mathcal{N}(0, \delta^2)$
      \For{\texttt{t=T:1}}
        \State $\bar{\epsilon} = F_{\Phi_{NN}}(x_t, v, t)$ 

        // \textit{Predict noise.}
        \State  $dx = [\theta_t(\mu-x)-g(t)^2 (-\frac{\bar\epsilon_t}{v_t})]\mathrm dt + g(t)d\hat{\omega}$

        // \textit{Approximation of score for Eq.~\ref{eq:deg_sde_forward}.}

        \State $x^*_{i-1} = \arg\min\limits_{x_{i-1}}[-\log p(x_{i-1}|x_i, x_0)]$

// \textit{Get the optimal reverse state minimizing the negative log-likelihood.
}    
      \EndFor
\Statex \textbf{RETURN} $x_0$
\end{algorithmic}
\end{algorithm}

\subsection{Terrain Prior Encoder}
\label{sec:TPE}
\begin{figure}[t]
    \centering
    \includegraphics[width=10cm]{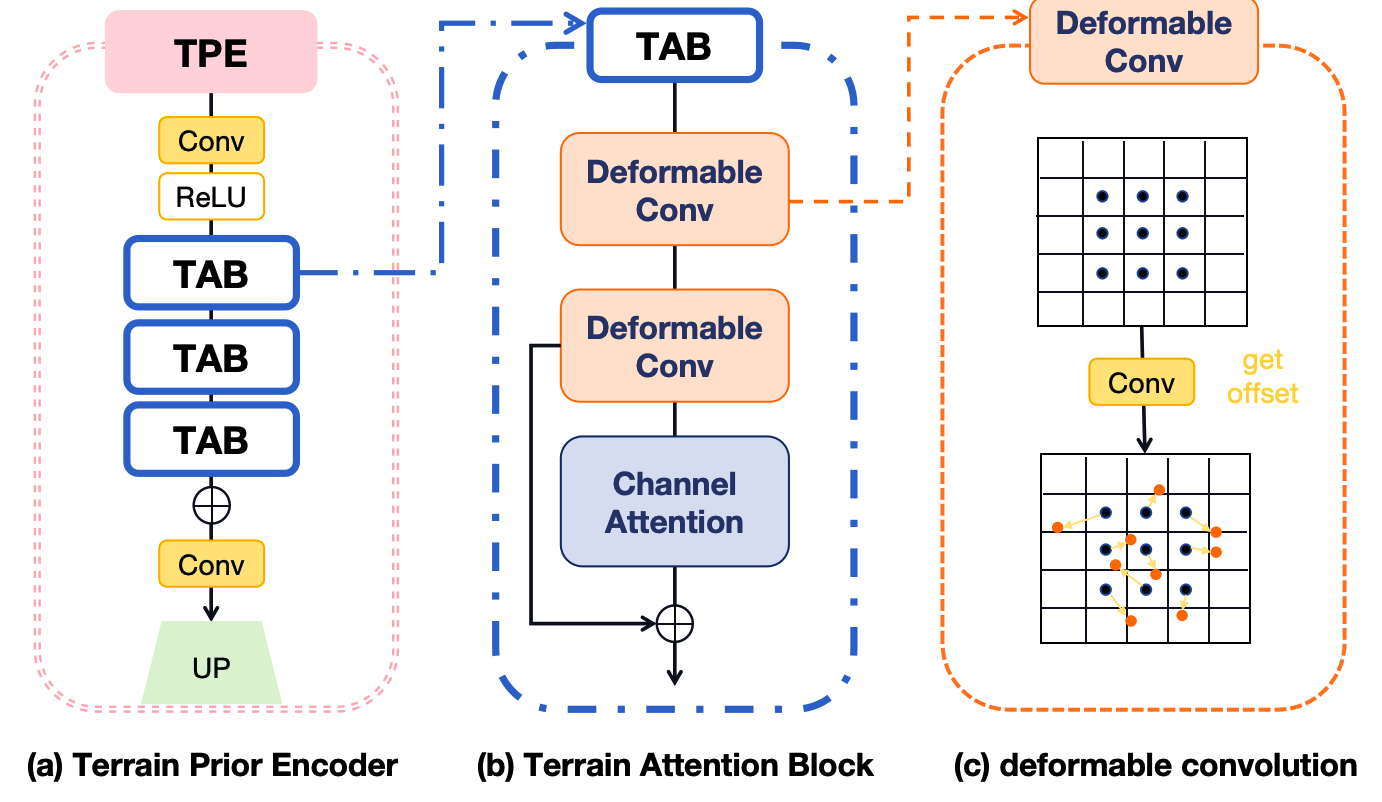}
    \caption{A TPE in (a) is composed of 3 TABs with details in (b). A deformable convolution in a TAB is illustrated in (c). }
    \label{fig:modules}
\end{figure}

The current diffusion pipelines used in computer vision tasks generally feed upsampled LR images to the noise predictor, which results in a lack of structural information, such as \citep{ir-sde,refusion}. EdiffSR \citep{ediffsr} goes a step further by utilizing the additional deep image prior of the LR image.
However, the mentioned approaches are not appropriate for capturing the structural details of terrain features.
%
%
Terrain exhibits significant irregularity across regions and patches within a dataset, particularly at ridges and saddles. Standard convolution layers extract features using regular weight kernels, limiting feature extraction capability.
Therefore, the TPE adopts deformable convolution as the basic operation.
As shown in Fig.~\ref{fig:modules} (a), the TPE consists of three Terrain Attention Blocks (TAB) in Fig.~\ref{fig:modules} (b). TAB encompasses two deformable convolutions shown in Fig.~\ref{fig:modules} (c), and a channel attention block \citep{cab} in the end.
The deformable convolutions incorporated offsets learned by the regular convolution to perform learnable-pattern sampling of locations.
\subsection{Noise Prediction Module}
\label{sec:eab}
\begin{figure}[H]
    \centering
    \includegraphics[width=14cm]{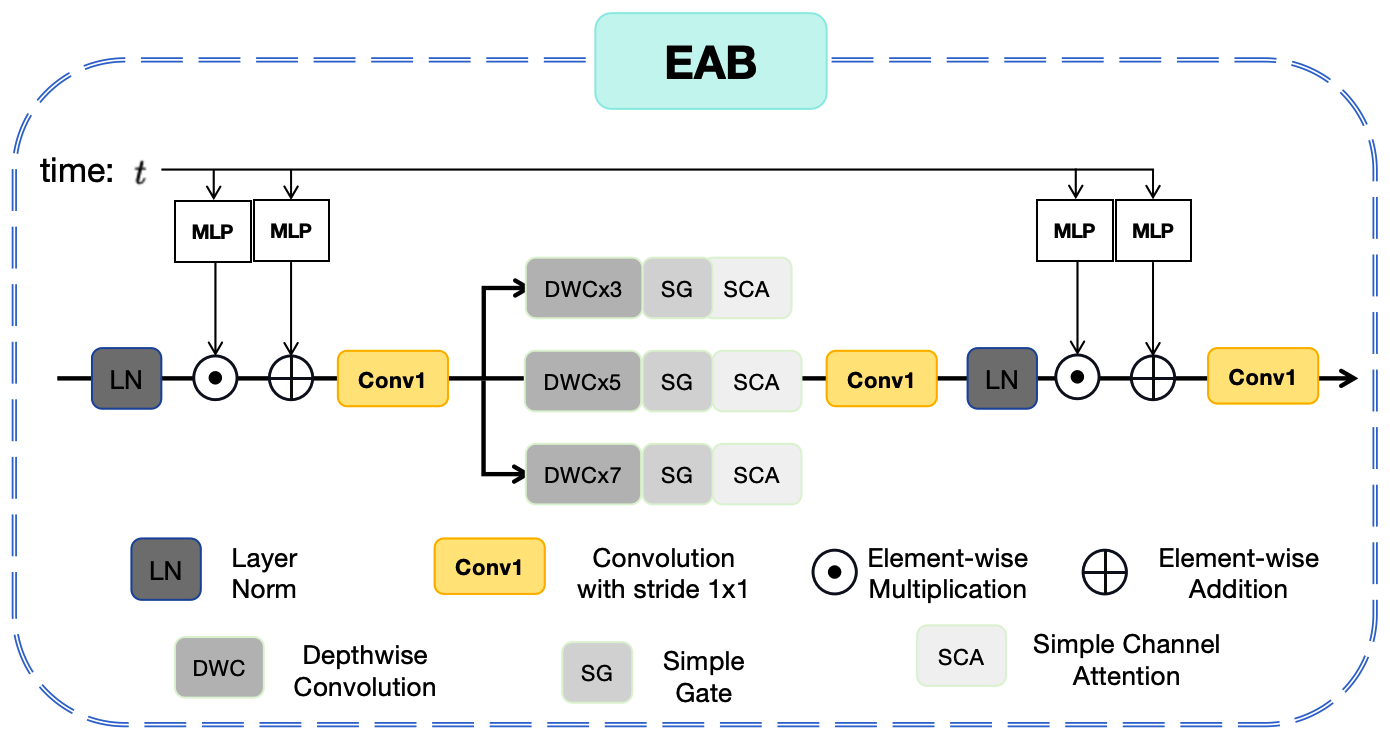}
    \caption{The Efficient Attention Block (EAB) of $F_{\Phi_{NN}}$ in the pipeline.}
    \label{fig:eab}
\end{figure}

To enable an fast and efficient forward process for DEM degradation, lightweight modules are assembled as Efficient Attention Blocks (EAB) adopted in the pipeline.
The detail of the repeated EABs in Fig.~\ref{fig:pipeline} is illustrated in Fig.~\ref{fig:eab}. 
Rather than using the standard U-Net, 
the noise predictor of ET-SDE incorporates simple channel attention (SCA), simple gate operation (SG) similar with \citep{refusion,ediffsr}, 
and the standard convolutions are replaced by depthwise convolution (DWC). 
Given the learned prior from the TPE module, the sampled time stamp $t$ is embedded within the EAB via MLPs to form the coefficients $\alpha$ and $\beta$, which modulate the input terrain prior $X_{TP}$. The process can be written as $X = \alpha \odot LN(X_{TP})$, where $LN$ indicates layer normalization.
Then $X$ is fed to a one-dimension convolution $F = conv_{1\times 1}(X)$.
To capture the feature at multiple scales, three parallel depthwise convolution blocks are applied to $F$, denoted as $DWCx3$, $DWCx5$ and $DWCx7$. Then the concatenated multiscale features are fed to the next layers, as shown in Fig.~\ref{fig:eab}.

\subsection{The Loss Function}
\label{sec:loss}
\paragraph{SDE Loss}
The basic and core optimization goal is the loss function of the SDE:
\begin{align}
    L_{sde} 
  &=\sum\limits_{i=1}^{T}\gamma_t \mathrm{E}[\|\underbrace{x(t) - (dx(t))_{\Tilde{\epsilon}_{\Phi}}}_{reversed\quad x(t-1)} - x^*(t-1)\|]\\ \nonumber
  &= \sum\limits_{t=0}^T\gamma_t\mathrm{E}[\|\hat{\epsilon}_{\Phi}(h_t, D_{LQ}, t) - \epsilon_t\|]
  \label{eq:loss_sde}
\end{align}
where $\gamma_t \in \{\gamma_1, \ldots, \gamma_T\}$ refers to the positive weight of step $t$, and $x^*(t-1)$ is the recursive optimal reversed state.




\paragraph{Gradient Loss}
To encourage the model to accurately model the ridges, gradient of two directions are utilized for topological supervision:
\begin{align}
  L_{edge} = \frac{1}{N}\sum_{i=1}^N((\frac{\partial \hat{h}}{\partial x} - \frac{\partial h_{gt}}{\partial x})^2 + (\frac{\partial\hat{h}}{\partial x} - \frac{\partial h_{gt}}{\partial y})^2)  
\end{align}
where $N$ is the number of points in the DEM involved in the computation.

\section{Experiment}
\label{sec:exp}
This section assesses the effectiveness of the proposed ET-SDE in three common refinements to improve DEM quality. These include super-resolution, void filling, and denoising. We compare ET-SDE with state-of-the-art methods in each respective field in Section~\ref{sec:sr} and Section~\ref{sec:voidfill}.
Section~\ref{sec:denoise} validates its effectiveness in removing various types of noise. 
Finally, Section~\ref{sec:ablation} conducts ablation studies on different modules.
%
\subsection{Basic Setup}
\subsubsection{Implementation Details}
The ET-SDE pipeline for the experiments incorporates 3 TABs in the TPE blocks to enhance the feature extraction with channels set to be 4.
The settings of the noise-predicting neural network follow the canonical diffusion pipeline \citep{ir-sde}. The internal channels of the convolutions are set to 64, the encoder contains 14, 1, 1, 1 EABs at each depth, and the decoder holds 1 EAB at each depth.
For the super-resolution task, for each batch with a batch size of 4500000 iterations are set for training. The initial learning rate is 4e5, with a cosine scheduler and an AdamW optimizer with $\beta_1=0.9, \beta_2=0.999$. The total diffusion step $T=50$ is set for the Pyrenees area with patch size of $96\times 96$, and $T=100$ is set for the Pyrenees area with patch size of $256\times 256$.
All the experiments are conducted with PyTorch on one NVIDIA RTX 3090 GPU with 24 GB memory. 
\subsubsection{Metrics}
To comprehensively evaluate the performance of super-resolution models, we use metrics from both a topological and an image perspective.
\begin{itemize}
    \item RMSE (Root Mean Square Error) of altitude: the standard deviation of the residuals between the ground truth and the estimated height map.
    \item PSNR (Peak Signal to Noise Ratio):
    The ratio between the maximum possible value (power) of a signal and the power of distorting noise affects the quality of its representation, manifesting the visual fidelity of the restored DEM.
    \item SSIM (Structural Similarity Index Measure): 
    The perceptual metric quantifies multiscale DEM quality degradation as an image, with various windows of the DEM patch. 
\end{itemize}
\subsubsection{Evaluation Area}
To evaluate the robustness and effectiveness of ET-SDE for super-resolution, we adopted two challenging mountainous datasets, one kept the same with EBCF-CDEM \citep{continuous}, with a relatively smaller patch size, and the other one with relatively larger DEM patches of $256\times 256$ with two super-resolution of 2 and 4 times of scales. The 
Both of the Pyrenees and the Mount Tai region are renowned for their complex terrain and diverse geomorphological features, \eg,  faults, folds, joints, and other structural forms. 
These structural features have had an important impact on the landforms and hydrological conditions,
making it suitable for testing point-based void completion methods. 

\begin{figure}
\centering
    \includegraphics[width=14cm]{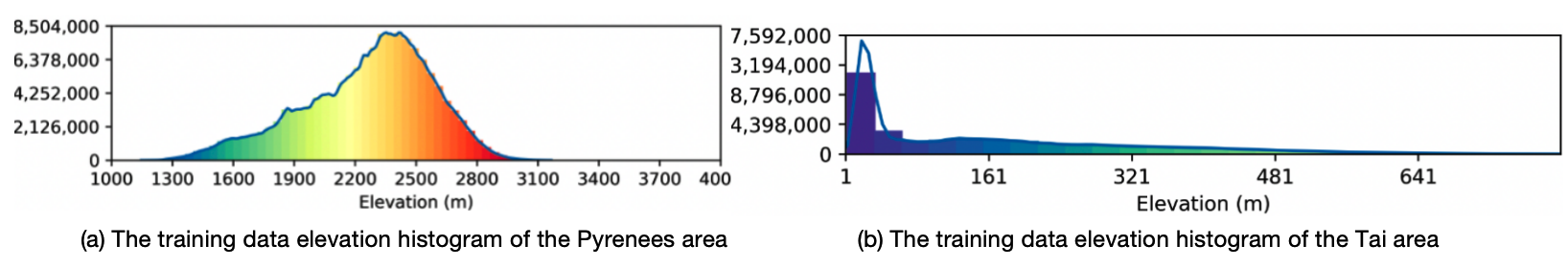}
    \caption{The elevation histograms of the Pyrenees area and the Tai area.
    }
    \label{fig:datahist}
\end{figure}
\paragraph{Dataset 1: Pyrenees}
The first study area is located around the Pyrenees mountain.
The DEM data of the Pyrenees has a resolution of 2$m$ and is divided into 10 regions, covering a total area of 643 square kilometers in the mountainous areas. The large-region DEMs are divided into non-overlapped small patches, each with a size of $96 \times 96$. $90\%$ of the DEM patches are randomly selected for training, while the remaining $10\%$ are used for testing. The Pyrenees area is used for the joint restoration task, with the resolution of the low-quality DEM to be 15$m$. 
\paragraph{Dataset 2: Mount Tai}
The second study area is located around Mount Tai, with elevation ascending abruptly.
%
The area is densely adorned with a profusion of springs, intricately interwoven with a network of rivers and streams, forming a unique hydrological landscape.
The elevation of the study area ranges from 3$m$ to 680$m$, with the histogram in Fig.~\ref{fig:datahist} (b).
The original data is obtained from the Advanced Spaceborne Thermal Emission and Reflection Radiometer (ASTER) \footnote{\url{https://gdemdl.aster.jspacesystems.or.jp/index_en.html}}.
The area is cropped into non-overlapped small patches of size $256\times 256$. $90\%$ of the small patches are split for training, and $10\%$ for testing.

\subsection{DEM super-resolution}
\label{sec:sr}
\subsubsection{Benchmarks}
To validate the effectiveness of the proposed ET-SDE on the DEM super-resolution task, 
we have chosen the EBCF-CDEM method to represent the deep-learning-based approach for its state-of-the-art performance over various types of methods. No other deep models were used in our study since EBCF-CDEM has reported the best performance. We have also chosen the bicubic interpolation as a representation of the traditional super-resolution method.
%
%
In the original implementation of EBCF-CDEM, the metrics are calculated in a cropped sub-patch with borders cut out. 
While For fairness, 
all the metrics are calculated for the whole patch. 
To evaluate the capability of ET-SDE to generalize across scales, the model is trained for 4x super-resolution, but tested for both 2x super-resolution and 4x super-resolution.
\subsubsection{Results of super resolution}
\paragraph{Results on Smaller Patches of Pyrenees}
Previous studies have commonly subdivided DEMs into smaller patches, primarily to accommodate methods that rely on localized information extraction using smaller windows \citep{continuous}. 
To ensure a fair comparison with the SOTA EBCF-CDEM, we maintain consistent data settings across our experiments. Detailed results of these comparative evaluations are presented in Table~\ref{tab:sr_pyrenees}.
\begin{table}[H]
\centering
\caption{Quantitative comparison of ET-SDE with smaller patches of the Pyrenees from the resolution of 15m to 2m.}
\begin{tabular}{c|ccc}
\toprule
model& Bicubic & EBCF-CDEM &ET-SDE\\
\midrule
PSNR & 28.57&\textbf{39.61}& 36.15 \\
SSIM & 0.72& \textbf{0.95}& 0.91\\
RMSE & 4.03& 2.72& \textbf{2.15}\\
\bottomrule
\end{tabular}
\label{tab:sr_pyrenees}
\end{table}
According to Table~\ref{tab:sr_pyrenees}, it is evident that ET-SDE demonstrates superior geometric performance in terms of 
RMSE on smaller DEM patches, although it performed slightly worse than the state-of-the-art model in terms of visual quality.
\paragraph{Results on Larger Patches of Mount Tai} 
In contrast to the small patches, this section investigates the capability of ET-SDE on larger DEM patches with more details and variations. Specifically, we examine the performance on $256 \times 256$ patches of the Mount Tai areas. The quantitative comparison of 4 times super-resolution is presented in Table~\ref{tab:sr_tai_aster}. This comparison provides a detailed analysis of the results for the larger DEM patches, allowing for a more comprehensive evaluation of the ET-SDE method.

\begin{figure}
\centering
    \includegraphics[width=14cm]{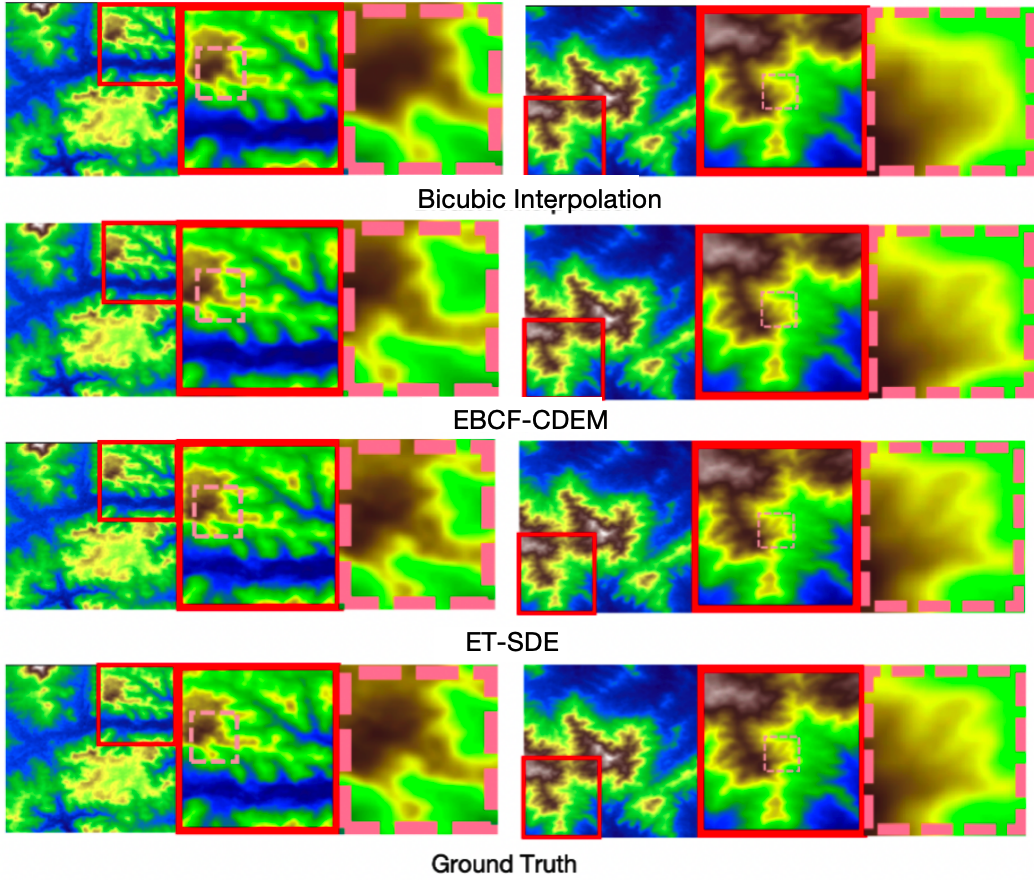}
    \caption{Qualitative comparison of ET-SDE with benchmarks of 2x super-resolution on Mount Tai area. 
    }
    \label{fig:results_asterx2}
\end{figure}

\begin{figure}[H]
\centering
    \includegraphics[width=14cm]{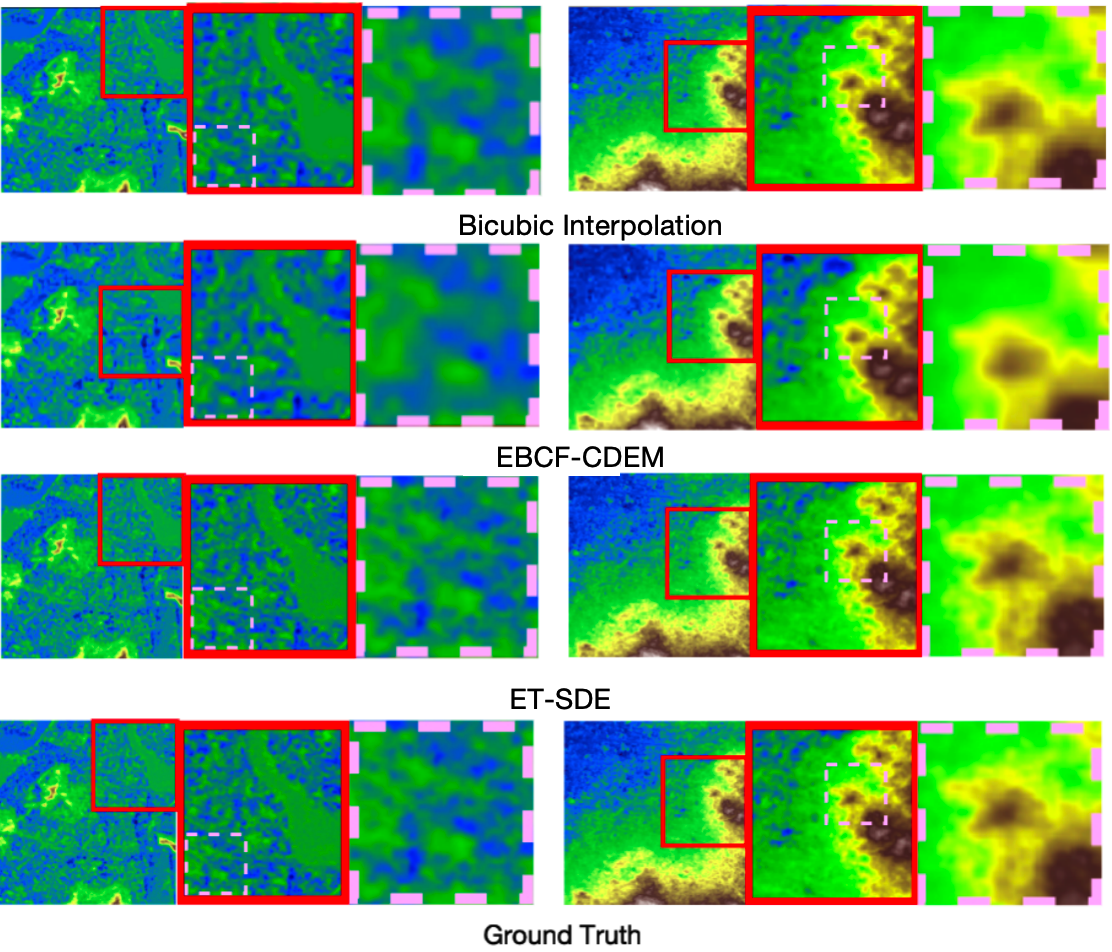}
    \caption{Qualitative comparison of ET-SDE with benchmarks of 4x super resolution on Mount Tai area. 
    }
    \label{fig:results_asterx4}
\end{figure}

\begin{table}[H]
\centering
\caption{Quantitative comparison of ET-SDE with on larger patches of the Mount Tai for Super-Resolution.}
\resizebox{0.9\textwidth}{!}{
\begin{tabular}{c|ccc|ccc}
\toprule
 &\multicolumn{3}{c|}{2x} & \multicolumn{3}{c}{4x} \\
 \hline
model& Bicubic & EBCF-CDEM &ET-SDE&Bicubic & EBCF-CDEM &ET-SDE\\
\midrule
PSNR &25.41&25.92 &\textbf{28.23} &22.39&21.44 &\textbf{26.51}\\
SSIM &0.69 & 0.72&\textbf{0.79} &0.65 & 0.67&\textbf{0.75} \\
RMSE &6.93&4.45&\textbf{3.82}&8.29&6.69&\textbf{4.13} \\
\bottomrule
\end{tabular}
}
\label{tab:sr_tai_aster}
\end{table}
While all evaluated models yield comprehensive results, the bicubic interpolation method is noted for its tendency to produce outputs that lack intricate details. 
In contrast, as Fig.~\ref{fig:results_asterx2} and Fig.~\ref{fig:results_asterx4} show, the EBCF-CDEM model exhibits an over-smoothing effect, which is primarily attributed to its reliance on locally continuous implicit representations. 
%
To sum up, the capability of each model varies concerning the scale of analysis.
EBCF-CDEM demonstrates superior performance in handling smaller patches of DEMs.
Conversely, ET-SDE showcases enhanced robustness and efficiency when processing larger patch sizes within the DEM framework. This differentiation underscores the strengths and limitations of each approach in addressing the spatial variability and resolution requirements inherent in DEM reconstruction and analysis.

\subsection{DEM void filling}
\label{sec:voidfill}
To evaluate ET-SDE on the void filling task, we select the state-of-the-art void filling method Diff-DEM \citep{diff-dem} based on a basic DDPM for comparison.
The bicubic interpolation is removed from the benchmarks because of its powerlessness.
In order to evaluate the model's adaptability and completion ability for different types of voids, we still adopt the two challenging study areas: the Pyrenees region and Mount Tai. 
The selected mountainous areas pose bigger challenges to the void-filling task, compared to the urban area introduced by Diff-DEM.

Diff-DEM assumes voids appearing in rectangular patches on DEMs of size $256\times 256$.
In addition, apart from the rectangular voids brought up by Diff-DEM, we also assume the existence of multiple random voids without regular shapes.
In the experimental design
we also introduce the DEMs with random voids on smaller DEMs of size $96\times 96$.

\subsubsection{Experimental Setup}
\paragraph{Benchmarks}

To effectively evaluate ET-SDE, we will utilize the state-of-the-art generative model Diff-DEM with various configurations. 
As mentioned in \citep{diff-dem}, better restoration performance is achieved with larger sampling steps during inference. For easier tasks, 128 steps are sufficient, while 512 steps are required for harder tasks.
Additionally, the inference process of Diff-DEM is dependent on the input void mask. However, in some cases, the void masks may not be available. 
Hence, we will compare ET-SDE under different sampling steps and conditions:
\begin{itemize}
    \item Diff-DEM 128: with 128 sampling steps at inference.
    \item Diff-DEM 256: with 256 sampling steps at inference.
    \item Diff-DEM wom: conditioning on the input low-quality DEM rather than the masks.
\end{itemize}

Interpolation is not considered this time because of its unavailability. 
\paragraph{Dataset Preprocessing}
To generate irregular voids on DEMs of smaller sizes, the random generation process starts from $N_{center}$ points randomly chosen from the DEM file, forming a $r_v\times r_v$ square of null values, and then $N_{center}$ random walk of $T_{walk}$ steps spread the null mask across the whole DEM mask.
The detailed summarization with notation can be referred to in Table~\ref{tab:masks_points}.

We first applied a point-based void generation method based on a random walk in the Pyrenees region to simulate small-scale data loss. Subsequently, we utilized a traditional patch-based void generation method in the Mount Tai area to simulate larger-scale data loss. By employing these two distinct types of voids, we can comprehensively assess the overall capability of the void completion model, including its ability to restore local details and maintain large-scale topographic features.

The second type of rectangular voids share the same settings with \citep{diff-dem} in Table~\ref{tab:masks_patches}.


\begin{minipage}[c]{0.5\textwidth}
   \captionof{table}{Notations of the void masks for the Pyrenees area.}
    \centering
    \begin{tabular}{c|ccc}
    \toprule
      Notation & $N_{center}$ & $T_{walk}$ &$r_v$    \\
      \midrule
       M-421  &4&2&1\\
       M-537  & 5&3&7\\
       M-623 & 6&2&3\\
       \bottomrule
    \end{tabular}
    \label{tab:masks_points}
\end{minipage}
\begin{minipage}[c]{0.4\textwidth}
    \captionof{table}{Notations of the rectangular void masks for the Mount Tai dataset.}
    \centering
    \begin{tabular}{c|cc}
    \toprule
      Notation & $w_{min}$ & $w_{mask}$ \\
      \midrule
V-64-96 &64 & 96\\
V-96-128& 96& 128\\
V-128-160 & 128& 160\\
       \bottomrule
    \end{tabular}
    \label{tab:masks_patches}
\end{minipage}

\subsubsection{Comparison of Random Voids}
\begin{figure}[H]
    \centering
    \includegraphics[width=0.95\linewidth]{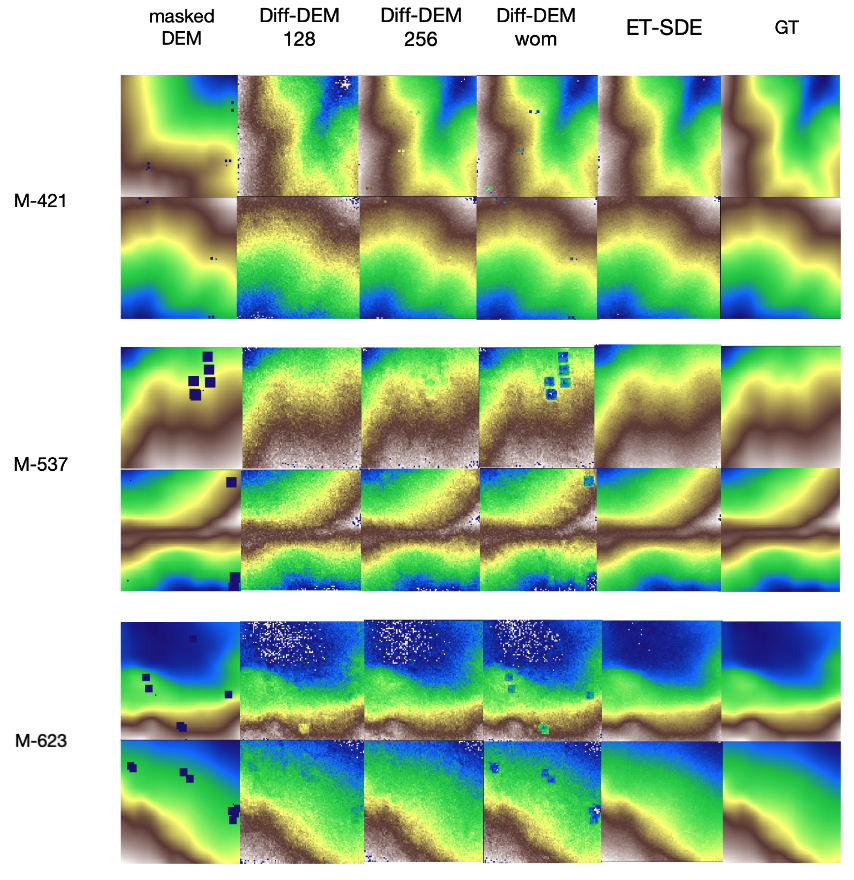}
    \caption{Comparison of the void filling results with the SOTA Diff-DEM with different inference steps.}
    \label{fig:pyrenees_void_filling}
\end{figure}

For random voids in the Pyrenees area, the performance comparison between ET-SDE and Diff-DEM with different settings is shown in Fig.~\ref{fig:pyrenees_void_filling}, and the quantitative comparison is given in Table~\ref{tab:pyrenees_void_filling}.
\begin{table}[H]
    \centering
    \caption{Quantitative comparison of DEM void filling for the Pyrenees area.}
    \resizebox{0.95\textwidth}{!}{
    \begin{tabular}{c|c|cccc}
    \toprule
        void&metric &Diff-DEM 128& Diff-DEM 256& Diff-DEM wom&ET-SDE  \\
         \midrule
         \multirow{3}{*}{M-421}&
         PSNR& 14.91&15.20&13.29& \textbf{23.31}\\
         &SSIM& 0.32 & 0.32 & 0.30 & \textbf{0.78}\\
         &RMSE& 8.84&8.69&9.22&\textbf{6.18}         \\
         \hline

                  \multirow{3}{*}{M-537}&
         PSNR& 14.88&14.25&12.05& \textbf{22.31}\\
         &SSIM& 0.30 & 0.31 & 0.25 & \textbf{0.69}\\
         &RMSE& 10.86&10.59&16.31&\textbf{6.73}         \\
         \hline
                 \multirow{3}{*}{M-623}&
         PSNR& 14.88&14.25&12.19& \textbf{21.76}\\
         &SSIM& 0.29 & 0.29 & 0.23 & \textbf{0.68}\\
         &RMSE& 11.37&11.61&15.97&\textbf{6.41}         \\
    \bottomrule
    \end{tabular}
}
\label{tab:pyrenees_void_filling}
\end{table}
 As demonstrated in Fig.~\ref{fig:pyrenees_void_filling}, augmenting the number of sampling steps indeed improves the overall performance of Diff-DEM, yet minor flaws persist. In scenarios where void masks are absent, noticeable edges of voids and sporadic noises are evident. Conversely, ET-SDE delivers commendable results without relying on void masks for direction. Broadly speaking, ET-SDE exhibits superior efficacy and robustness in comparison to Diff-DEM.

\subsubsection{Comparison of Rectangular Voids}
\begin{figure}[H]
    \centering
    \includegraphics[width=0.95\linewidth]{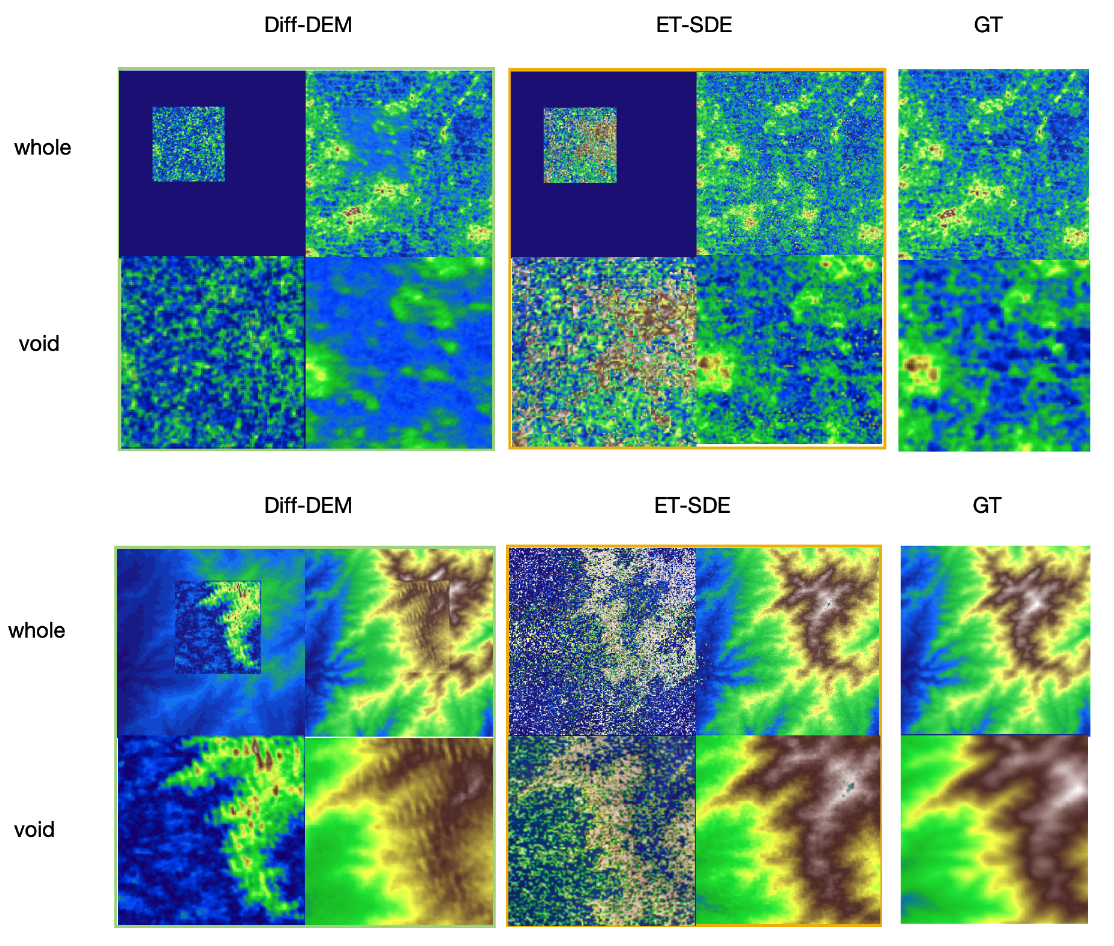}
    \caption{Comparison of the void filling results with the SOTA Diff-DEM with different inference steps.}
    \label{fig:aster_void_filling}
\end{figure}

For rectangular voids in the mount Tai area, the sampling steps of Diff-DEM are set to be 300 to adapt to the $256\times 256$ patches.
This section is dedicated to evaluating the performance of ET-SDE against the SOTA techniques on DEMs, taking into account varying levels of complexity determined by the size of the void sizes.
The models are trained on DEMs across different sizes of rectangular voids.
the performance comparison between ET-SDE and Diff-DEM is shown in Fig.~\ref{fig:aster_void_filling}, and the quantitative comparison is given in Table~\ref{tab:aster_void_filling}.
\begin{table}[H]
    \centering
    \caption{Quantitative comparison of DEM void filling for the mount Tai Area with different void sizes.}
    \resizebox{0.9\textwidth}{!}{
    \begin{tabular}{c|cccccc}
    \toprule
   & \multicolumn{2}{c}{$64\times 64$} &
    \multicolumn{2}{c}{$64\times 96$} &
    \multicolumn{2}{c}{$96\times 128$} \\
    \hline
         &Diff-DEM &ET-SDE &Diff-DEM &ET-SDE&Diff-DEM &ET-SDE \\
         \hline
         PSNR& 15.62&\textbf{24.21} & 14.92& \textbf{22.97}& 14.37 & \textbf{21.81} \\
         SSIM& 0.47&\textbf{0.81}&0.43&\textbf{0.77}&0.42&\textbf{0.73}\\
         RMSE&8.97&\textbf{7.02}&9.21&\textbf{7.36}& 12.04&\textbf{7.94}\\
         \bottomrule
    \end{tabular}
}
    \label{tab:aster_void_filling}
\end{table}
As shown in Fig.~\ref{fig:aster_void_filling}, Diff-DEM tends to learn and transfer the whole DEM structure to generate the void, since the diffusion process is conditioned on the entire DEM rather than understanding the common patterns of the area.
The phenomenon becomes more apparent when the void size increases, while ET-SDE yields more robust results with finer details of the joining edges. 
\subsection{DEM Denoising}
\label{sec:denoise}
This section examines the effectiveness of ET-SDE in denoising DEM data. Since there are few existing methods specifically tailored for denoising DEM, we also consider the added noises as irregular voids and compare them with Diff-DEM for reference.
\subsubsection{Experimental Setup}
\paragraph{Benchmarks}
As a representation of the advanced generative model, Diff-DEM shows great capability in void-filling in urban areas. 
Since the added noises cover the whole DEM, and larger sampling steps yield better reconstruction results, we set 10x larger sampling steps for the denoising tasks for Diff-DEM, \ie, Diff-DEM 1000 for 1000 steps and Diff-DEM 3000 for 3000 steps.
Besides, to evaluate the generalization ability of ET-SDE between tasks, we also apply the ET-SDE model originally trained for void-filling on the denoising task.
\paragraph{Dataset Preprocessing}
We select three typical noises to evaluate ET-SDE, the Poisson noise, speckle noise, and pepper noise.
The training and testing splits are kept the same with the super-resolution task.

\subsubsection{Comparison of DEM Denoising}
\begin{figure}
    \centering
\includegraphics[width=0.95\linewidth]{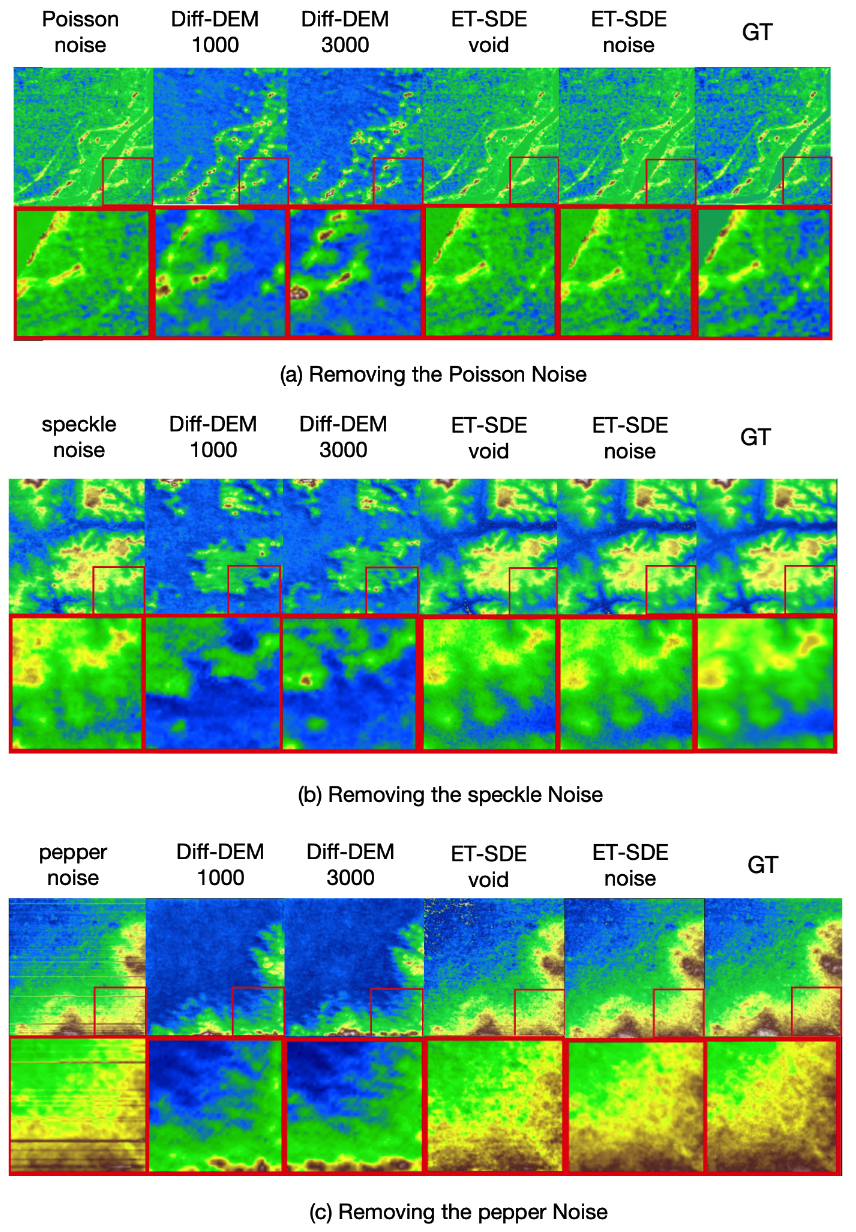}
    \caption{Comparion of denoising of the Mount Tai area.}
    \label{fig:aster_noise}
\end{figure}

Fig.~\ref{fig:aster_noise} offers a visual qualitative comparison of the noise reduction tasks. 
The individual qualitative analyses for each noise type—Poisson, speckle, and pepper—are listed in Tables~\ref{tab:aster_poisson}, \ref{tab:aster_speckle}, and \ref{tab:aster_pepper}, respectively.
\begin{table}[H]
    \centering
    \caption{Quantitative comparison of DEM denoising of Poisson noise.}
    \resizebox{0.9\textwidth}{!}{
    \begin{tabular}{c|cccc}
    \toprule
         &Diff-DEM 1000& Diff-DEM 3000& ET-SDE noise&ET-SDE void \\
         \midrule
         PSNR& 11.73&11.59&\textbf{22.94}& 21.86\\
         SSIM& 0.24 & 0.22 & 0.72 & \textbf{0.78}\\
         RMSE& 14.37& 14.22& \textbf{6.31}& 6.54        \\
    \bottomrule
    \end{tabular}
    }
    \label{tab:aster_poisson}
\end{table}
\begin{table}[H]
    \centering
    \caption{Quantitative comparison of DEM denoising of speckle noise.}
    \resizebox{0.9\textwidth}{!}{
    \begin{tabular}{c|cccc}
    \toprule
         &Diff-DEM 1000& Diff-DEM 3000& ET-SDE noise&ET-SDE void \\
         \midrule
         PSNR& 11.98&12.03&\textbf{23.62}& 23.09\\
         SSIM& 0.29 & 0.28 & 0.68 & \textbf{0.70}\\
         RMSE&15.94  &15.28 &\textbf{7.13} & 7.17       \\
    \bottomrule
    \end{tabular}
    }

    \label{tab:aster_speckle}
\end{table}
\begin{table}[H]
    \centering
        \caption{Quantitative comparison of DEM denoising of pepper noise.}
    \resizebox{0.9\textwidth}{!}{
    \begin{tabular}{c|cccc}
    \toprule
         &Diff-DEM 1000& Diff-DEM 3000& ET-SDE noise&ET-SDE void \\
         \midrule
         PSNR& 12.82&12.85&19.59&\textbf{20.32} \\
         SSIM& 0.24 & 0.24 & 0.59&\textbf{0.68} \\
         RMSE & 13.62 & 12.475 & 9.77&\textbf{8.53} \\
    \bottomrule
    \end{tabular}
    }
    \label{tab:aster_pepper}
\end{table}

As can be seen from Fig.~\ref{fig:aster_noise}, adding noise leads the estimation of Diff-DEM to smaller values than the ground truth.
With no explicit void masks provided, Diff-DEM can not restore the real terrain information, even though the total structure has been kept.
Meanwhile, ET-SDE shows robustness without mask guidance and provides sound restorations.
Notably, the ET-SDE trained for void-filling appears to show similar performance for the inference of the denoising task, even compared to the ET-SDE trained for denoising.
\subsection{Model Efficiency}
To manifest the efficiency and lightweight of the proposed ET-SDE, four metrics are considered, including the trainable parameters Param, Floating Point Operations Per Second FLOPs, and inference time $Time$.
FLOPs-96 represents the FLOPs with input shape of $96\times 96$, FLOPs-256 for $256\times 256$, Time-96 for $96\times 96$, and Time-256 for $256\times 256$.

The comparison of inference time and the model parameters are shown in Table~\ref{tab:efficiency}.
\begin{table}[H]
    \centering
        \caption{Model efficiency comparison with state-of-the-art models on DEMs of size $96\times 96$.}
        \resizebox{0.9\textwidth}{!}{
    \begin{tabular}{c|ccccc}
    \toprule
        sites & \textbf{Param}$\downarrow$& \textbf{FLOPs-96}$\downarrow$& \textbf{FLOPs-256}$\downarrow$&\textbf{Time-96}$\downarrow$&\textbf{Time-256}$\downarrow$\\
        \midrule
EBCF-CDEM&\textbf{1.43M}&90.08G&1702.18G&26.23s&77.55s\\
\hline
DIF-DEM &56.44M & \textbf{7.49G}& 212.90G& 38.59s&114.79s\\
\hline
ET-SDE& 30.75M&8.53G&\textbf{202.23G} & \textbf{5.04s}&\textbf{7.95s}\\
         \bottomrule
    \end{tabular}
}
    \label{tab:efficiency}
\end{table}

In Table~\ref{tab:efficiency}, it is evident that ET-SDE significantly reduces the inference speed compared to the state-of-the-art works. It is important to note that the FLOPs and times are calculated for the entire DEM for EBCF-CDEM rather than for each patch, and the patch sizes are maintained at their original settings. 
For DIF-DEM, although the total FLOPs are smaller than ET-SDE for smaller DEMs, the independent sampling steps consume a significant amount of time. In summary, ET-SDE demonstrates the fastest inference speed with a smaller computation cost for larger input.





\subsection{Ablation Study}
\label{sec:ablation}
To investigate the utility of the Terrain Prior Encoder (TPE) and the Efficient Attention Block (EAB), we compare the baseline ET-SDE with changing settings:
\begin{itemize}
    \item \textbf{wo}-TPE: the baseline ET-SDE without a prior encoder.
    \item \textbf{wo}-EAB: the baseline ET-SDE with vanilla attention module. 
    \item \textbf{wo}-TPE-EAB: the baseline ET-SDE without a prior encoder, and with the vanilla attention module. 
\end{itemize}
The ablation study, focusing on the 4x super-resolution task of Mount Tai, assesses performance based on both reconstruction quality and computational efficiency. 
\begin{table}[H]
    \centering
    \caption{The ablation study of the model design.}
\resizebox{0.9\textwidth}{!}{
\begin{tabular}{c|ccccccc}
\toprule
model & TPE&EAB&PSNR  & SSIM  &RMSE & Param&FLOPs\\
\midrule
  \textbf{wo}-TPE &\XSolidBrush & \CheckmarkBold & 23.79 &0.67&4.82 & 30.26M&6.48G\\
\textbf{wo}-EAB & \CheckmarkBold & \CheckmarkBold &24.25 & 0.71 & 4.66 &137.13M &37.99G \\
  \textbf{wo}-TPE-EAB &\XSolidBrush &\XSolidBrush &24.10& 0.69&5.35&137.62M&40.04G\\
  ET-SDE  & \CheckmarkBold & \CheckmarkBold & 26.51&0.75&4.13&30.75M&8.53G
  \\
\bottomrule
\end{tabular}
}
    \label{tab:ablation_TPE}
\end{table}
The ablation comparison is presented in Table~\ref{tab:ablation_TPE}. It is evident that using prior encoding leads to superior super-resolution results without significantly increasing computation costs. Besides, EAB helps a lot in reducing total parameters and computation.
In summary, incorporating TPE and EAB proves beneficial for enhancing overall performance.

   

\section{Conclusion}
\label{sec:conclusion}
This paper proposes a multipurpose generative diffusing pipeline based on the stochastic differential equation to restore DEM with low-resolution voids.
Different from the previous methods of restoring DEM with different defects separately, ET-SDE can handle multiple problems and can be applied to relatively larger DEM patches.
ET-SDE is conditioned on deep terrain prior to image super-resolution pipelines, and the implementation incorporates more efficient modules.
ET-SDE does not refer to local grids for neighboring information, thus it is more robust to larger voids, and larger input sizes.
Experiments have confirmed the effectiveness of the ET-SDE for super-resolution, with or without varying levels of missing data. Additionally, the ET-SDE is suitable for processing larger input patches.
Despite of its robustness for sparse, small voids and patchified voids, ET-SDE can not handle multiple large voids.
In the future, a more robust deep terrain prior could be designed, to enhance the capability.
Pretraining across different datasets is also a promising derivative.

\section*{Acknowledgement}
This work was supported in part by the Special Funds for Creative Research Groups under Grant 2022C61540, and  Laboratory of China-ASEAN Satellite Remote Sensing Applications, Ministry of Natural Resources of the People's Republic of China (ZDMY202302).

\bibliographystyle{elsarticle-harv} 
\bibliography{ref}

\begin{thebibliography}{50}
\expandafter\ifx\csname natexlab\endcsname\relax\def\natexlab#1{#1}\fi
\providecommand{\url}[1]{\texttt{#1}}
\providecommand{\href}[2]{#2}
\providecommand{\path}[1]{#1}
\providecommand{\DOIprefix}{doi:}
\providecommand{\ArXivprefix}{arXiv:}
\providecommand{\URLprefix}{URL: }
\providecommand{\Pubmedprefix}{pmid:}
\providecommand{\doi}[1]{\href{http://dx.doi.org/#1}{\path{#1}}}
\providecommand{\Pubmed}[1]{\href{pmid:#1}{\path{#1}}}
\providecommand{\bibinfo}[2]{#2}
\ifx\xfnm\relax \def\xfnm[#1]{\unskip,\space#1}\fi
\bibitem[{Ajvazi and Czimber(2019)}]{comp-inter19}
\bibinfo{author}{Ajvazi, B.}, \bibinfo{author}{Czimber, K.}, \bibinfo{year}{2019}.
\newblock \bibinfo{title}{A comparative analysis of different dem interpolation methods in gis: case study of rahovec, kosovo}.
\newblock \bibinfo{journal}{Geodesy and cartography} \bibinfo{volume}{45}, \bibinfo{pages}{43--48}.
\bibitem[{Beutel et~al.(2010)Beutel, M{\o}lhave and Agarwal}]{naturalneighbor}
\bibinfo{author}{Beutel, A.}, \bibinfo{author}{M{\o}lhave, T.}, \bibinfo{author}{Agarwal, P.K.}, \bibinfo{year}{2010}.
\newblock \bibinfo{title}{Natural neighbor interpolation based grid dem construction using a gpu}, in: \bibinfo{booktitle}{Proceedings of the 18th SIGSPATIAL International Conference on Advances in geographic information Systems}, pp. \bibinfo{pages}{172--181}.
\bibitem[{Ceylan and Gutmann(2018)}]{NCSNs}
\bibinfo{author}{Ceylan, C.}, \bibinfo{author}{Gutmann, M.U.}, \bibinfo{year}{2018}.
\newblock \bibinfo{title}{Conditional noise-contrastive estimation of unnormalised models}, in: \bibinfo{booktitle}{International Conference on Machine Learning}, \bibinfo{organization}{PMLR}. pp. \bibinfo{pages}{726--734}.
\bibitem[{Chen et~al.(2021)Chen, Liu and Wang}]{liif}
\bibinfo{author}{Chen, Y.}, \bibinfo{author}{Liu, S.}, \bibinfo{author}{Wang, X.}, \bibinfo{year}{2021}.
\newblock \bibinfo{title}{Learning continuous image representation with local implicit image function}, in: \bibinfo{booktitle}{Proceedings of the IEEE/CVF conference on computer vision and pattern recognition}, pp. \bibinfo{pages}{8628--8638}.
\bibitem[{Chen et~al.(2016)Chen, Wang, Xu and Hou}]{cnn1st_dem}
\bibinfo{author}{Chen, Z.}, \bibinfo{author}{Wang, X.}, \bibinfo{author}{Xu, Z.}, \bibinfo{author}{Hou, W.}, \bibinfo{year}{2016}.
\newblock \bibinfo{title}{Convolutional neural network based dem super resolution}.
\newblock \bibinfo{journal}{The International Archives of the Photogrammetry, Remote Sensing and Spatial Information Sciences} \bibinfo{volume}{41}, \bibinfo{pages}{247--250}.
\bibitem[{Choi et~al.(2021)Choi, Kim, Jeong, Gwon and Yoon}]{ilvr}
\bibinfo{author}{Choi, J.}, \bibinfo{author}{Kim, S.}, \bibinfo{author}{Jeong, Y.}, \bibinfo{author}{Gwon, Y.}, \bibinfo{author}{Yoon, S.}, \bibinfo{year}{2021}.
\newblock \bibinfo{title}{Ilvr: Conditioning method for denoising diffusion probabilistic models}.
\newblock \bibinfo{journal}{arXiv preprint arXiv:2108.02938} .
\bibitem[{Croitoru et~al.(2023)Croitoru, Hondru, Ionescu and Shah}]{diffusion_review}
\bibinfo{author}{Croitoru, F.A.}, \bibinfo{author}{Hondru, V.}, \bibinfo{author}{Ionescu, R.T.}, \bibinfo{author}{Shah, M.}, \bibinfo{year}{2023}.
\newblock \bibinfo{title}{Diffusion models in vision: A survey}.
\newblock \bibinfo{journal}{IEEE Transactions on Pattern Analysis and Machine Intelligence} .
\bibitem[{Demiray et~al.(2021)Demiray, Sit and Demir}]{d-srgan}
\bibinfo{author}{Demiray, B.Z.}, \bibinfo{author}{Sit, M.}, \bibinfo{author}{Demir, I.}, \bibinfo{year}{2021}.
\newblock \bibinfo{title}{D-srgan: Dem super-resolution with generative adversarial networks}.
\newblock \bibinfo{journal}{SN Computer Science} \bibinfo{volume}{2}, \bibinfo{pages}{1--11}.
\bibitem[{Farr et~al.(2007)Farr, Rosen, Caro, Crippen, Duren, Hensley, Kobrick, Paller, Rodriguez, Roth et~al.}]{srtm}
\bibinfo{author}{Farr, T.G.}, \bibinfo{author}{Rosen, P.A.}, \bibinfo{author}{Caro, E.}, \bibinfo{author}{Crippen, R.}, \bibinfo{author}{Duren, R.}, \bibinfo{author}{Hensley, S.}, \bibinfo{author}{Kobrick, M.}, \bibinfo{author}{Paller, M.}, \bibinfo{author}{Rodriguez, E.}, \bibinfo{author}{Roth, L.}, et~al., \bibinfo{year}{2007}.
\newblock \bibinfo{title}{The shuttle radar topography mission}.
\newblock \bibinfo{journal}{Reviews of geophysics} \bibinfo{volume}{45}.
\bibitem[{Gavriil et~al.(2019)Gavriil, Muntingh and Barrowclough}]{dem_void_gan}
\bibinfo{author}{Gavriil, K.}, \bibinfo{author}{Muntingh, G.}, \bibinfo{author}{Barrowclough, O.J.}, \bibinfo{year}{2019}.
\newblock \bibinfo{title}{Void filling of digital elevation models with deep generative models}.
\newblock \bibinfo{journal}{IEEE Geoscience and Remote Sensing Letters} \bibinfo{volume}{16}, \bibinfo{pages}{1645--1649}.
\bibitem[{Ho et~al.(2020)Ho, Jain and Abbeel}]{ddpm}
\bibinfo{author}{Ho, J.}, \bibinfo{author}{Jain, A.}, \bibinfo{author}{Abbeel, P.}, \bibinfo{year}{2020}.
\newblock \bibinfo{title}{Denoising diffusion probabilistic models}.
\newblock \bibinfo{journal}{Advances in neural information processing systems} \bibinfo{volume}{33}, \bibinfo{pages}{6840--6851}.
\bibitem[{Jiang et~al.(2020)Jiang, Hu, Xia, Liang, Soltoggio and Kabir}]{multiscale}
\bibinfo{author}{Jiang, L.}, \bibinfo{author}{Hu, Y.}, \bibinfo{author}{Xia, X.}, \bibinfo{author}{Liang, Q.}, \bibinfo{author}{Soltoggio, A.}, \bibinfo{author}{Kabir, S.R.}, \bibinfo{year}{2020}.
\newblock \bibinfo{title}{A multi-scale mapping approach based on a deep learning cnn model for reconstructing high-resolution urban dems}.
\newblock \bibinfo{journal}{Water} \bibinfo{volume}{12}, \bibinfo{pages}{1369}.
\bibitem[{Kakavas and Nikolakopoulos(2021)}]{dem_landslides}
\bibinfo{author}{Kakavas, M.P.}, \bibinfo{author}{Nikolakopoulos, K.G.}, \bibinfo{year}{2021}.
\newblock \bibinfo{title}{Digital elevation models of rockfalls and landslides: a review and meta-analysis}.
\newblock \bibinfo{journal}{Geosciences} \bibinfo{volume}{11}, \bibinfo{pages}{256}.
\bibitem[{Krieger et~al.(2007)Krieger, Moreira, Fiedler, Hajnsek, Werner, Younis and Zink}]{tandem}
\bibinfo{author}{Krieger, G.}, \bibinfo{author}{Moreira, A.}, \bibinfo{author}{Fiedler, H.}, \bibinfo{author}{Hajnsek, I.}, \bibinfo{author}{Werner, M.}, \bibinfo{author}{Younis, M.}, \bibinfo{author}{Zink, M.}, \bibinfo{year}{2007}.
\newblock \bibinfo{title}{Tandem-x: A satellite formation for high-resolution sar interferometry}.
\newblock \bibinfo{journal}{IEEE Transactions on Geoscience and Remote Sensing} \bibinfo{volume}{45}, \bibinfo{pages}{3317--3341}.
\bibitem[{Kulp and Strauss(2018)}]{coastaldem}
\bibinfo{author}{Kulp, S.A.}, \bibinfo{author}{Strauss, B.H.}, \bibinfo{year}{2018}.
\newblock \bibinfo{title}{Coastaldem: A global coastal digital elevation model improved from srtm using a neural network}.
\newblock \bibinfo{journal}{Remote sensing of environment} \bibinfo{volume}{206}, \bibinfo{pages}{231--239}.
\bibitem[{Lepcha et~al.(2023)Lepcha, Goyal, Dogra and Goyal}]{sisr_review}
\bibinfo{author}{Lepcha, D.C.}, \bibinfo{author}{Goyal, B.}, \bibinfo{author}{Dogra, A.}, \bibinfo{author}{Goyal, V.}, \bibinfo{year}{2023}.
\newblock \bibinfo{title}{Image super-resolution: A comprehensive review, recent trends, challenges and applications}.
\newblock \bibinfo{journal}{Information Fusion} \bibinfo{volume}{91}, \bibinfo{pages}{230--260}.
\bibitem[{Li et~al.(2022a)Li, Hu, Cheng, Xiong, Tang and Strobl}]{ridgeline}
\bibinfo{author}{Li, S.}, \bibinfo{author}{Hu, G.}, \bibinfo{author}{Cheng, X.}, \bibinfo{author}{Xiong, L.}, \bibinfo{author}{Tang, G.}, \bibinfo{author}{Strobl, J.}, \bibinfo{year}{2022}a.
\newblock \bibinfo{title}{Integrating topographic knowledge into deep learning for the void-filling of digital elevation models}.
\newblock \bibinfo{journal}{Remote Sensing of Environment} \bibinfo{volume}{269}, \bibinfo{pages}{112818}.
\bibitem[{Li et~al.(2022b)Li, Hu, Cheng, Xiong, Tang and Strobl}]{TKCGAN}
\bibinfo{author}{Li, S.}, \bibinfo{author}{Hu, G.}, \bibinfo{author}{Cheng, X.}, \bibinfo{author}{Xiong, L.}, \bibinfo{author}{Tang, G.}, \bibinfo{author}{Strobl, J.}, \bibinfo{year}{2022}b.
\newblock \bibinfo{title}{Integrating topographic knowledge into deep learning for the void-filling of digital elevation models}.
\newblock \bibinfo{journal}{Remote Sensing of Environment} \bibinfo{volume}{269}, \bibinfo{pages}{112818}.
\bibitem[{Li et~al.(2023)Li, Ren, Jin, Lan, Wang, Zeng, Wang and Chen}]{diffusion_ir_review}
\bibinfo{author}{Li, X.}, \bibinfo{author}{Ren, Y.}, \bibinfo{author}{Jin, X.}, \bibinfo{author}{Lan, C.}, \bibinfo{author}{Wang, X.}, \bibinfo{author}{Zeng, W.}, \bibinfo{author}{Wang, X.}, \bibinfo{author}{Chen, Z.}, \bibinfo{year}{2023}.
\newblock \bibinfo{title}{Diffusion models for image restoration and enhancement--a comprehensive survey}.
\newblock \bibinfo{journal}{arXiv preprint arXiv:2308.09388} .
\bibitem[{Lim et~al.(2017)Lim, Son, Kim, Nah and Mu~Lee}]{edsr}
\bibinfo{author}{Lim, B.}, \bibinfo{author}{Son, S.}, \bibinfo{author}{Kim, H.}, \bibinfo{author}{Nah, S.}, \bibinfo{author}{Mu~Lee, K.}, \bibinfo{year}{2017}.
\newblock \bibinfo{title}{Enhanced deep residual networks for single image super-resolution}, in: \bibinfo{booktitle}{Proceedings of the IEEE conference on computer vision and pattern recognition workshops}, pp. \bibinfo{pages}{136--144}.
\bibitem[{Lo and Peters(2024)}]{diff-dem}
\bibinfo{author}{Lo, K.S.H.}, \bibinfo{author}{Peters, J.}, \bibinfo{year}{2024}.
\newblock \bibinfo{title}{Diff-dem: A diffusion probabilistic approach to digital elevation model void filling}.
\newblock \bibinfo{journal}{IEEE Geoscience and Remote Sensing Letters} .
\bibitem[{Luo et~al.(2023a)Luo, Gustafsson, Zhao, Sj{\"o}lund and Sch{\"o}n}]{ir-sde}
\bibinfo{author}{Luo, Z.}, \bibinfo{author}{Gustafsson, F.K.}, \bibinfo{author}{Zhao, Z.}, \bibinfo{author}{Sj{\"o}lund, J.}, \bibinfo{author}{Sch{\"o}n, T.B.}, \bibinfo{year}{2023}a.
\newblock \bibinfo{title}{Image restoration with mean-reverting stochastic differential equations}.
\newblock \bibinfo{journal}{arXiv preprint arXiv:2301.11699} .
\bibitem[{Luo et~al.(2023b)Luo, Gustafsson, Zhao, Sj{\"o}lund and Sch{\"o}n}]{refusion}
\bibinfo{author}{Luo, Z.}, \bibinfo{author}{Gustafsson, F.K.}, \bibinfo{author}{Zhao, Z.}, \bibinfo{author}{Sj{\"o}lund, J.}, \bibinfo{author}{Sch{\"o}n, T.B.}, \bibinfo{year}{2023}b.
\newblock \bibinfo{title}{Refusion: Enabling large-size realistic image restoration with latent-space diffusion models}, in: \bibinfo{booktitle}{Proceedings of the IEEE/CVF Conference on Computer Vision and Pattern Recognition}, pp. \bibinfo{pages}{1680--1691}.
\bibitem[{Mesa-Mingorance and Ariza-L{\'o}pez(2020)}]{dem_accuracy_review}
\bibinfo{author}{Mesa-Mingorance, J.L.}, \bibinfo{author}{Ariza-L{\'o}pez, F.J.}, \bibinfo{year}{2020}.
\newblock \bibinfo{title}{Accuracy assessment of digital elevation models (dems): A critical review of practices of the past three decades}.
\newblock \bibinfo{journal}{Remote Sensing} \bibinfo{volume}{12}, \bibinfo{pages}{2630}.
\bibitem[{O'Loughlin et~al.(2016)O'Loughlin, Paiva, Durand, Alsdorf and Bates}]{o2016multi}
\bibinfo{author}{O'Loughlin, F.E.}, \bibinfo{author}{Paiva, R.C.}, \bibinfo{author}{Durand, M.}, \bibinfo{author}{Alsdorf, D.}, \bibinfo{author}{Bates, P.}, \bibinfo{year}{2016}.
\newblock \bibinfo{title}{A multi-sensor approach towards a global vegetation corrected srtm dem product}.
\newblock \bibinfo{journal}{Remote Sensing of Environment} \bibinfo{volume}{182}, \bibinfo{pages}{49--59}.
\bibitem[{Pavlova(2017)}]{comp_inter17}
\bibinfo{author}{Pavlova, A.}, \bibinfo{year}{2017}.
\newblock \bibinfo{title}{Analysis of elevation interpolation methods for creating digital elevation models}.
\newblock \bibinfo{journal}{Optoelectronics, Instrumentation and Data Processing} \bibinfo{volume}{53}, \bibinfo{pages}{171--177}.
\bibitem[{Polidori and El~Hage(2020)}]{dem_assess_review}
\bibinfo{author}{Polidori, L.}, \bibinfo{author}{El~Hage, M.}, \bibinfo{year}{2020}.
\newblock \bibinfo{title}{Digital elevation model quality assessment methods: A critical review}.
\newblock \bibinfo{journal}{Remote sensing} \bibinfo{volume}{12}, \bibinfo{pages}{3522}.
\bibitem[{Roostaee and Deng(2020)}]{hydrology}
\bibinfo{author}{Roostaee, M.}, \bibinfo{author}{Deng, Z.}, \bibinfo{year}{2020}.
\newblock \bibinfo{title}{Effects of digital elevation model resolution on watershed-based hydrologic simulation}.
\newblock \bibinfo{journal}{Water Resources Management} \bibinfo{volume}{34}, \bibinfo{pages}{2433--2447}.
\bibitem[{Ruiz-Lend{\'\i}nez et~al.(2023)Ruiz-Lend{\'\i}nez, Ariza-L{\'o}pez, Reinoso-Gordo, Ure{\~n}a-C{\'a}mara and Quesada-Real}]{sota_dem_review}
\bibinfo{author}{Ruiz-Lend{\'\i}nez, J.J.}, \bibinfo{author}{Ariza-L{\'o}pez, F.J.}, \bibinfo{author}{Reinoso-Gordo, J.F.}, \bibinfo{author}{Ure{\~n}a-C{\'a}mara, M.A.}, \bibinfo{author}{Quesada-Real, F.J.}, \bibinfo{year}{2023}.
\newblock \bibinfo{title}{Deep learning methods applied to digital elevation models: state of the art}.
\newblock \bibinfo{journal}{Geocarto International} \bibinfo{volume}{38}, \bibinfo{pages}{2252389}.
\bibitem[{Saharia et~al.(2022a)Saharia, Chan, Chang, Lee, Ho, Salimans, Fleet and Norouzi}]{palette}
\bibinfo{author}{Saharia, C.}, \bibinfo{author}{Chan, W.}, \bibinfo{author}{Chang, H.}, \bibinfo{author}{Lee, C.}, \bibinfo{author}{Ho, J.}, \bibinfo{author}{Salimans, T.}, \bibinfo{author}{Fleet, D.}, \bibinfo{author}{Norouzi, M.}, \bibinfo{year}{2022}a.
\newblock \bibinfo{title}{Palette: Image-to-image diffusion models}, in: \bibinfo{booktitle}{ACM SIGGRAPH 2022 Conference Proceedings}, pp. \bibinfo{pages}{1--10}.
\bibitem[{Saharia et~al.(2022b)Saharia, Ho, Chan, Salimans, Fleet and Norouzi}]{sr3}
\bibinfo{author}{Saharia, C.}, \bibinfo{author}{Ho, J.}, \bibinfo{author}{Chan, W.}, \bibinfo{author}{Salimans, T.}, \bibinfo{author}{Fleet, D.J.}, \bibinfo{author}{Norouzi, M.}, \bibinfo{year}{2022}b.
\newblock \bibinfo{title}{Image super-resolution via iterative refinement}.
\newblock \bibinfo{journal}{IEEE Transactions on Pattern Analysis and Machine Intelligence} \bibinfo{volume}{45}, \bibinfo{pages}{4713--4726}.
\bibitem[{Saied et~al.(2020)Saied, Elshafey and Mahmoud}]{sar-speckle-dem}
\bibinfo{author}{Saied, S.K.}, \bibinfo{author}{Elshafey, M.A.}, \bibinfo{author}{Mahmoud, T.A.}, \bibinfo{year}{2020}.
\newblock \bibinfo{title}{Digital elevation model enhancement using cnn-based despeckled sar images}, in: \bibinfo{booktitle}{2020 IEEE Aerospace Conference}, \bibinfo{organization}{IEEE}. pp. \bibinfo{pages}{1--8}.
\bibitem[{Sankaran and Holmes(2023)}]{gan_review}
\bibinfo{author}{Sankaran, K.}, \bibinfo{author}{Holmes, S.P.}, \bibinfo{year}{2023}.
\newblock \bibinfo{title}{Generative models: An interdisciplinary perspective}.
\newblock \bibinfo{journal}{Annual Review of Statistics and Its Application} \bibinfo{volume}{10}, \bibinfo{pages}{325--352}.
\bibitem[{Setianto and Triandini(2013)}]{IDW}
\bibinfo{author}{Setianto, A.}, \bibinfo{author}{Triandini, T.}, \bibinfo{year}{2013}.
\newblock \bibinfo{title}{Comparison of kriging and inverse distance weighted (idw) interpolation methods in lineament extraction and analysis}.
\newblock \bibinfo{journal}{Journal of Applied Geology} \bibinfo{volume}{5}.
\bibitem[{Song et~al.(2020a)Song, Sohl-Dickstein, Kingma, Kumar, Ermon and Poole}]{score-sde}
\bibinfo{author}{Song, Y.}, \bibinfo{author}{Sohl-Dickstein, J.}, \bibinfo{author}{Kingma, D.P.}, \bibinfo{author}{Kumar, A.}, \bibinfo{author}{Ermon, S.}, \bibinfo{author}{Poole, B.}, \bibinfo{year}{2020}a.
\newblock \bibinfo{title}{Score-based generative modeling through stochastic differential equations}.
\newblock \bibinfo{journal}{arXiv preprint arXiv:2011.13456} .
\bibitem[{Song et~al.(2020b)Song, Sohl-Dickstein, Kingma, Kumar, Ermon and Poole}]{sde}
\bibinfo{author}{Song, Y.}, \bibinfo{author}{Sohl-Dickstein, J.}, \bibinfo{author}{Kingma, D.P.}, \bibinfo{author}{Kumar, A.}, \bibinfo{author}{Ermon, S.}, \bibinfo{author}{Poole, B.}, \bibinfo{year}{2020}b.
\newblock \bibinfo{title}{Score-based generative modeling through stochastic differential equations}.
\newblock \bibinfo{journal}{arXiv preprint arXiv:2011.13456} .
\bibitem[{Toutin(2008)}]{aster}
\bibinfo{author}{Toutin, T.}, \bibinfo{year}{2008}.
\newblock \bibinfo{title}{Aster dems for geomatic and geoscientific applications: a review}.
\newblock \bibinfo{journal}{International Journal of Remote Sensing} \bibinfo{volume}{29}, \bibinfo{pages}{1855--1875}.
\bibitem[{Van~Beers and Kleijnen(2004)}]{kriging}
\bibinfo{author}{Van~Beers, W.C.}, \bibinfo{author}{Kleijnen, J.P.}, \bibinfo{year}{2004}.
\newblock \bibinfo{title}{Kriging interpolation in simulation: a survey}, in: \bibinfo{booktitle}{Proceedings of the 2004 Winter Simulation Conference, 2004.}, \bibinfo{organization}{IEEE}.
\bibitem[{Xiao et~al.(2023)Xiao, Yuan, Jiang, He, Jin and Zhang}]{ediffsr}
\bibinfo{author}{Xiao, Y.}, \bibinfo{author}{Yuan, Q.}, \bibinfo{author}{Jiang, K.}, \bibinfo{author}{He, J.}, \bibinfo{author}{Jin, X.}, \bibinfo{author}{Zhang, L.}, \bibinfo{year}{2023}.
\newblock \bibinfo{title}{Ediffsr: An efficient diffusion probabilistic model for remote sensing image super-resolution}.
\newblock \bibinfo{journal}{IEEE Transactions on Geoscience and Remote Sensing} .
\bibitem[{Xu et~al.(2019)Xu, Chen, Yi, Gui, Hou and Ding}]{gradient}
\bibinfo{author}{Xu, Z.}, \bibinfo{author}{Chen, Z.}, \bibinfo{author}{Yi, W.}, \bibinfo{author}{Gui, Q.}, \bibinfo{author}{Hou, W.}, \bibinfo{author}{Ding, M.}, \bibinfo{year}{2019}.
\newblock \bibinfo{title}{Deep gradient prior network for dem super-resolution: Transfer learning from image to dem}.
\newblock \bibinfo{journal}{ISPRS Journal of Photogrammetry and Remote Sensing} \bibinfo{volume}{150}, \bibinfo{pages}{80--90}.
\bibitem[{Xu et~al.(2015)Xu, Wang, Chen, Xiong, Ding and Hou}]{xu2015nonlocal}
\bibinfo{author}{Xu, Z.}, \bibinfo{author}{Wang, X.}, \bibinfo{author}{Chen, Z.}, \bibinfo{author}{Xiong, D.}, \bibinfo{author}{Ding, M.}, \bibinfo{author}{Hou, W.}, \bibinfo{year}{2015}.
\newblock \bibinfo{title}{Nonlocal similarity based dem super resolution}.
\newblock \bibinfo{journal}{ISPRS Journal of Photogrammetry and Remote Sensing} \bibinfo{volume}{110}, \bibinfo{pages}{48--54}.
\bibitem[{Yan et~al.(2021)Yan, Tang and Zhang}]{gan_interpolation2}
\bibinfo{author}{Yan, L.}, \bibinfo{author}{Tang, X.}, \bibinfo{author}{Zhang, Y.}, \bibinfo{year}{2021}.
\newblock \bibinfo{title}{High accuracy interpolation of dem using generative adversarial network}.
\newblock \bibinfo{journal}{Remote Sensing} \bibinfo{volume}{13}, \bibinfo{pages}{676}.
\bibitem[{Yang et~al.(2023)Yang, Xu, Lv, Zhou, Zhu and Cheng}]{dem_terrain}
\bibinfo{author}{Yang, J.}, \bibinfo{author}{Xu, J.}, \bibinfo{author}{Lv, Y.}, \bibinfo{author}{Zhou, C.}, \bibinfo{author}{Zhu, Y.}, \bibinfo{author}{Cheng, W.}, \bibinfo{year}{2023}.
\newblock \bibinfo{title}{Deep learning-based automated terrain classification using high-resolution dem data}.
\newblock \bibinfo{journal}{International Journal of Applied Earth Observation and Geoinformation} \bibinfo{volume}{118}, \bibinfo{pages}{103249}.
\bibitem[{Yang et~al.(2011)Yang, Meng and Zhang}]{srtm_app}
\bibinfo{author}{Yang, L.}, \bibinfo{author}{Meng, X.}, \bibinfo{author}{Zhang, X.}, \bibinfo{year}{2011}.
\newblock \bibinfo{title}{Srtm dem and its application advances}.
\newblock \bibinfo{journal}{International Journal of Remote Sensing} \bibinfo{volume}{32}, \bibinfo{pages}{3875--3896}.
\bibitem[{Yao et~al.(2024)Yao, Cheng, Yang and Mozerov}]{continuous}
\bibinfo{author}{Yao, S.}, \bibinfo{author}{Cheng, Y.}, \bibinfo{author}{Yang, F.}, \bibinfo{author}{Mozerov, M.G.}, \bibinfo{year}{2024}.
\newblock \bibinfo{title}{A continuous digital elevation representation model for dem super-resolution}.
\newblock \bibinfo{journal}{ISPRS Journal of Photogrammetry and Remote Sensing} \bibinfo{volume}{208}, \bibinfo{pages}{1--13}.
\bibitem[{Zhang et~al.(2021)Zhang, Bian and Li}]{rspcn}
\bibinfo{author}{Zhang, R.}, \bibinfo{author}{Bian, S.}, \bibinfo{author}{Li, H.}, \bibinfo{year}{2021}.
\newblock \bibinfo{title}{Rspcn: super-resolution of digital elevation model based on recursive sub-pixel convolutional neural networks}.
\newblock \bibinfo{journal}{ISPRS International Journal of Geo-Information} \bibinfo{volume}{10}, \bibinfo{pages}{501}.
\bibitem[{Zhang et~al.(2018)Zhang, Li, Li, Wang, Zhong and Fu}]{cab}
\bibinfo{author}{Zhang, Y.}, \bibinfo{author}{Li, K.}, \bibinfo{author}{Li, K.}, \bibinfo{author}{Wang, L.}, \bibinfo{author}{Zhong, B.}, \bibinfo{author}{Fu, Y.}, \bibinfo{year}{2018}.
\newblock \bibinfo{title}{Image super-resolution using very deep residual channel attention networks}, in: \bibinfo{booktitle}{Proceedings of the European conference on computer vision (ECCV)}, pp. \bibinfo{pages}{286--301}.
\bibitem[{Zhang et~al.(2022)Zhang, Yu and Zhu}]{terrainfeatureaware}
\bibinfo{author}{Zhang, Y.}, \bibinfo{author}{Yu, W.}, \bibinfo{author}{Zhu, D.}, \bibinfo{year}{2022}.
\newblock \bibinfo{title}{Terrain feature-aware deep learning network for digital elevation model superresolution}.
\newblock \bibinfo{journal}{ISPRS Journal of Photogrammetry and Remote Sensing} \bibinfo{volume}{189}, \bibinfo{pages}{143--162}.
\bibitem[{Zhou et~al.(2022)Zhou, Song, Liang, Xu and Yue}]{dem_void_attgan}
\bibinfo{author}{Zhou, G.}, \bibinfo{author}{Song, B.}, \bibinfo{author}{Liang, P.}, \bibinfo{author}{Xu, J.}, \bibinfo{author}{Yue, T.}, \bibinfo{year}{2022}.
\newblock \bibinfo{title}{Voids filling of dem with multiattention generative adversarial network model}.
\newblock \bibinfo{journal}{Remote Sensing} \bibinfo{volume}{14}, \bibinfo{pages}{1206}.
\bibitem[{Zhu et~al.(2020)Zhu, Cheng, Zhang, Yao, Gao and Liu}]{gan_interpolation}
\bibinfo{author}{Zhu, D.}, \bibinfo{author}{Cheng, X.}, \bibinfo{author}{Zhang, F.}, \bibinfo{author}{Yao, X.}, \bibinfo{author}{Gao, Y.}, \bibinfo{author}{Liu, Y.}, \bibinfo{year}{2020}.
\newblock \bibinfo{title}{Spatial interpolation using conditional generative adversarial neural networks}.
\newblock \bibinfo{journal}{International Journal of Geographical Information Science} \bibinfo{volume}{34}, \bibinfo{pages}{735--758}.

\end{thebibliography}

\end{document}